\documentclass[a4paper,useAMS,usenatbib]{mn2e}

\usepackage{times}
\usepackage[dvips]{graphicx}

\usepackage[totalwidth=504pt, totalheight=680pt, letterpaper]{geometry}

\newcommand{\kms}{\ensuremath{\hbox{km}\cdot \hbox{s}^{-1}}}
\newcommand{\hmpc}{\ensuremath{h^{-1}\,\hbox{Mpc}}}

\newcommand{\etal}{{et al.}~}

\newcommand{\bfx}{{\bf x}}
\newcommand{\bfr}{{\bf r}}

\newcommand{\bfy}{{\bf y}}
\newcommand{\bfk}{{\bf k}}
\newcommand{\bfv}{{\bf v}}

\newcommand{\bfg}{{\bf g}}

\newcommand{\calF}{{\cal F}}
\newcommand{\calO}{{\cal O}}

\newcommand{\calR}{\mathcal{R}}
\newcommand{\eps}{{\epsilon}}
\newcommand{\bc}{\begin{center}}
\newcommand{\be}{\begin{equation}}
\newcommand{\ee}{\end{equation}}
\newcommand{\ec}{\end{center}}
\newcommand{\lan}{\langle}
\newcommand{\ran}{\rangle}

\newcommand{\de}{\delta}
\newcommand{\te}{\vartheta}

\newcommand{\err}{{\rho}}
\newcommand{\sig}{\sigma}

\newcommand{\spose}[1]{\hbox to 0pt{#1\hss}}
\newcommand{\lta}{\mathrel{\spose{\lower 3pt\hbox{$\mathchar"218$}}
 \raise 2.0pt\hbox{$\mathchar"13C$}}}
\newcommand{\gta}{\mathrel{\spose{\lower 3pt\hbox{$\mathchar"218$}}
 \raise 2.0pt\hbox{$\mathchar"13E$}}}

\title[2MASS dipole]{Towards the optimal window for the 2MASS dipole} 

\author[Chodorowski, Coiffard, Bilicki, Colombi \& Cieciel\c{a}g]
{Micha{\l} J.\ Chodorowski,$^1$\thanks{E-mail: michal@camk.edu.pl} 
Jean-Baptiste Coiffard,$^2$ Maciej Bilicki,$^1$
\newauthor St\'ephane Colombi$^3$ and Pawe{\l} Cieciel\c{a}g$^1$\\ 
$^1$Copernicus Astronomical Center, Bartycka 18, 00--716 Warsaw, 
Poland \\
$^2$Universit\'e Paris XI, 91400 Orsay, France \\
$^3$ Institut d'Astrophysique de Paris, CNRS, 98 bis Boulevard Arago, 
			75014 Paris, France}

\begin{document}
\maketitle


\begin{abstract}
A comparison of the 2MASS flux dipole to the CMB dipole can serve as a
method to constrain a combination of the cosmological parameter
$\Omega_m$ and the luminosity bias of the 2MASS survey. For this
constraint to be as tight as possible, it is necessary to maximize the
correlation between the two dipoles. This can be achieved by
optimizing the survey window through which the flux dipole is
measured. Here we explicitly construct such a window for the 2MASS
survey. The optimization in essence reduces to excluding from the
calculation of the flux dipole galaxies brighter than some limiting
magnitude $K_{\rm min}$ of the near-infrared $K_s$ band.  This
exclusion mitigates nonlinear effects and shot noise from small
scales, which decorrelate the 2MASS dipole from the CMB dipole. Under
the assumption of negligible shot noise we find that the optimal value
of $K_{\rm min}$ is about five. Inclusion of shot noise shifts the
optimal $K_{\rm min}$ to larger values. We present an analytical
formula for shot noise for the 2MASS flux dipole, to be used in
follow-up work with 2MASS data.

The misalignment angle between the two dipoles is a sensitive measure
of their correlation: the higher the correlation, the smaller the
expectation value of the angle. A minimum of the misalignment is thus
a sign of the optimal gravity window. We model analytically the
distribution function for the misalignment angle and show that the
misalignment estimated by Maller et al.\ is consistent with the
assumed underlying model (though it is greater than the expectation
value). We predict with about 90\% confidence that the misalignment
will decrease if 2MASS galaxies brighter than $K_{\rm min} = 5~\rm
mag$ are excluded from the calculation of the flux dipole. This
prediction has been indirectly confirmed by the results of Erdo{\u
g}du et al. The measured misalignment constitutes thus an alternative
way of finding the optimal value of $K_{\rm min}$: the latter
corresponds to a minimum of the former.
\end{abstract}

\section{introduction}
\label{sec:intro}
The dipole anisotropy of the cosmic microwave background (CMB) is
interpreted as a direct measure, via the Doppler shift, of the motion
of the Local Group (LG) relative to the CMB rest frame. The components
of this motion of non-cosmological origin (the motion of the Sun in
the Milky Way and the motion of the Milky Way in the LG) are known and
can be subtracted (e.g., Courteau \& van den Bergh 1999). When
transformed to the barycenter of the LG, the motion is towards $(l,b)
= (273^{\circ} \pm 3^{\circ},29^{\circ} \pm 3^{\circ})$, and of
amplitude $v_{\scriptscriptstyle \rm LG} = 627 \pm 22$ \kms, as
inferred from the first-year WMAP data (Bennett \etal 2003).

The kinematic interpretation of the CMB dipole is strongly supported
by its remarkable alignment with the dipole component of the
large-scale galaxy distribution (often called the `clustering
dipole'), inferred from various all-sky surveys. In the gravitational
instability scenario, this alignment is expected: peculiar velocities
of galaxies are induced gravitationally and are thus strongly coupled
to the large-scale matter distribution. Linear theory predicts the
peculiar velocity of the LG, $\bfv$, to be proportional to the LG
peculiar acceleration, caused by the gravitational pull of surrounding
matter inhomogeneities. Let us denote by $\de$ the mass density
contrast, $\de \equiv \varrho/\varrho_b - 1$, where $\varrho$ is the
mass density of matter and $\varrho_b$ is its average value. The
clustering dipole,
\be
\bfg\equiv\int\frac{{\rm d}^{3}r}{4\pi}\delta(\bfr)\frac{\bfr}{r^3}\,,
\label{eq:clust}
\ee
is a quantity {\em proportional\/} to the peculiar gravitational
acceleration (so we will call it interchangeably `scaled gravity'),
and can be estimated from a three-dimensional all-sky galaxy
survey. In the linear regime, the relation between the velocity and
the scaled gravity is
\be 
\bfv = H_0 f(\Omega_m) \bfg \,.
\label{eq:v-g}
\ee
Here, $H_0$ is the Hubble constant, $\Omega_{m}$ is the cosmic matter
density parameter and $f(\Omega_m) \simeq \Omega_m^{0.6}$ (e.g.,
Peebles 1980). For a spherical survey $\int {\rm d}^{3} r \: \bfr/r^3
= 0$, hence we can write
\be 
\bfg = \int\frac{{\rm d}^{3}r}{4\pi} \bigl[1 + \delta(\bfr)\bigr]
\frac{\bfr}{r^3} = \varrho_b^{-1}\!\! \int\frac{{\rm d}^{3}r}{4\pi}
\varrho(\bfr) \frac{\hat\bfr}{r^2} \,.
\label{eq:grav}
\ee 
In the following we will assume that dark matter (DM) in the Universe
is entirely locked in DM halos of luminous galaxies. 
Modelling galaxies as point particles, the observed density field is 
$\varrho(\bfr) = \sum_i m_i \de_D(\bfr - \bfr_i)$, where $\de_D$ is
Dirac's delta; $m_i$ and $\bfr_i$ are respectively the mass and
the position of the $i$-th galaxy. Substituting this equation into
Equation~(\ref{eq:grav}) yields for the scaled gravity
\be 
\bfg = \varrho_b^{-1} \sum_i \frac{m_i}{4 \pi} \frac{\hat\bfr_i}{r_i^2} \,;
\label{eq:grav_sum}
\ee
thus we see that the true gravitational acceleration equals to $4 \pi
G\varrho_b \bfg$. We will assume further that `light traces mass', or
that the mass-to-light ratio for galaxies is a universal constant,
$\Upsilon$. Then we can write
\be 
\bfg = \varrho_b^{-1} \sum_i \frac{\Upsilon L_i}{4 \pi}
\frac{\hat\bfr_i}{r_i^2} = \frac{\Upsilon}{\varrho_b} \sum_i S_i 
\hat\bfr_i \,.
\label{eq:grav_flux}
\ee
Here, $L_i$ is the luminosity of $i$-th galaxy and $S_i$ is its
observed flux, $S_i = L_i/4 \pi r_i^2$. In other words, since both the
gravity and the flux fall off as distance squared, the gravitational
acceleration of the LG is proportional to the dipole of the light
distribution (i.e., the {\em flux\/} dipole) for a constant
mass-to-light ratio. The sum in Equation~(\ref{eq:grav_flux}) is in
principle over all galaxies in the Universe, while in practice we have
at our disposal only finite, usually flux-limited, catalogs of
galaxies. In such catalogs, lower-mass dark matter halos will be
underrepresented by the survey galaxies. To account for this, we write
\be 
\bfg = \frac{\Upsilon}{\varrho_b b_L} \sum_{i=1}^{N} S_i 
\hat\bfr_i \,,
\label{eq:grav_bias}
\ee 
where $b_L$ is the resulting luminosity bias and $N$ is the total
number of galaxies in a given survey. Combining
Equation~(\ref{eq:v-g}) with Equation~(\ref{eq:grav_bias}) we obtain
finally (Erdo{\u g}du \etal 2006; hereafter E06)
\be
\bfv = \frac{H_0 \Omega_m^{0.6}}{\varrho_L b_L} \sum_{i=1}^{N} S_i 
\hat\bfr_i \,.
\label{eq:vel_fin}
\ee In the above we have used the fact that the mass-to-light ratio
$\Upsilon = \varrho_b / \varrho_L$, where $\varrho_L$ is the
luminosity density of the Universe. Equation~(\ref{eq:vel_fin}) shows
that in the linear theory one can predict the LG peculiar velocity
using solely an {\em angular\/} (two-dimensional) all-sky survey,
bypassing the lack of radial information, i.e.\
distances. Specifically, a comparison between the CMB dipole and the
flux dipole of a given survey can yield an estimate of the parameter
$\beta \equiv \Omega_m^{0.6}/b_L$.

Such a comparison was first performed by Yahil, Sandage \& Tamman
(1980) using the revised Shapley-Ames catalogue and by Davis \& Huchra
(1982) using the CfA catalogue, leading to the estimates of the flux
dipoles that were within $30^\circ$ from the CMB dipole. The inclusion
of redshift information, usage of progressively larger redshift
surveys and theoretical improvements of the analyses led to smaller
measured values of the misalignment. In particular, using the IRAS
$1.2$ Jy survey, Strauss \etal (1992, hereafter S92) found that the
clustering dipole points around $25^\circ$ away from the CMB
dipole. Using the further completed IRAS PSCz survey, Schmoldt et
al. 1999 (hereafter S99) obtained the clustering dipole within
$15^\circ$ of the CMB dipole. A similar analysis, based also on the
IRAS PSCz survey, performed by Rowan-Robinson et al. (2000),
determined the misalignment angle to be around $13^\circ$.

Two most recent analyses of the clustering dipole employed the Two
Micron All Sky Survey (2MASS; Skrutskie \etal 1997). In particular, to
compute the flux dipole, Maller \etal (2003; hereafter M03) used the
angular 2MASS extended source catalogue, with a limiting magnitude of
$K_s = 13.57$. (Approximately 740,000 galaxies covering $90$\% of the
sky.) E06 used the Two Micron All Sky {\sl Redshift\/} Survey (2MRS):
approximately 23,200 2MASS galaxies with measured redshifts, selected
from a total sample of about 24,800 galaxies with
(extinction-corrected) magnitudes smaller than $K_s = 11.25$.

2MASS is the first near-infrared ($J$$H$$K_s$ passbands) all-sky
survey. While most passbands tend to be sensitive to the instantaneous
star formation rate, $K_s$ passband is most sensitive to total stellar
mass (Bell \& de Jong 2001; Bell \etal 2003), making this band a
better tracer of total mass. 2MASS has an effective image resolution
of 1'' and a hundred times greater sensitivity than the far-infrared
{\it IRAS} survey. The photometric uniformity of the 2MASS survey is
better than $4$ per cent over the entire sky including the celestial
poles (e.g., Jarrett \etal 2003). The median depth of the survey is
$220$ \hmpc\ (Bell \etal 2003), a distance past where the clustering
dipole has been shown to converge.\footnote{The inclusion of galaxy
redshifts in the dipole analyses allowed the estimation of the
convergence depth, i.e.\ the distance at which most of the clustering
dipole is generated. There is a controversy whether this convergence
depth is about $50$ \hmpc, or rather $200$ \hmpc\ (for details see
E06). In either case, 2MASS is deep enough to provide a reliable
estimate of the clustering dipole. (But see Basilakos \& Plionis
2006.)}

Given all these advantages of 2MASS over other all-sky galaxy surveys,
it is perhaps surprising that the misalignment between the CMB dipole
and the 2MASS flux dipole is not smaller than the corresponding one
for {\em IRAS\/} galaxies. The value obtained by M03 is
$16^\circ$. For the 2MRS flux dipole, E06 obtained approximately
$21^\circ$.\footnote{E06 computed two kinds of the clustering
dipole. The second one, the number dipole, was even more misaligned
with the CMB dipole.} In this paper we aim at answering the following
questions. First: do we understand fully the origin of this
misalignment? Second: can one do better with 2MASS, and if so, how?

The answer to these questions is essential for optimal estimation
of the parameter $\beta$ by comparing the CMB dipole to the 2MASS
dipole. The stronger is the correlation between the two dipoles, the
smaller are statistical errors of such an estimate. Therefore, the
observational window, through which the 2MASS dipole is measured,
should be adapted to obtain the best correlation possible. The
misalignment angle is a sensitive measure of this correlation: the
higher the correlation, the smaller the angle. In other words, a
minimum of the misalignment angle is a sign of the optimal 2MASS
window. In this paper we will formally prove these statements. First,
we will derive the 2MASS window. Next, we will optimize it under the
assumption of negligible shot noise. Finally, we will demonstrate that
a minimum of the expectation value of the angle corresponds to minimal
variance of the resulting estimate of $\beta$.

Let us enumerate possible sources of the misalignment between the CMB
dipole and an all-sky galaxy survey flux dipole.
\begin{itemize}
\item $M/L \ne \rm const$. The constant mass-to-light ratio is
  probably a good assumption for (almost) all galaxies when averaged
  over many galaxies of the same luminosity. For individual galaxies,
  however, $M/L$ is expected to have some scatter. On the other hand,
  as stated earlier, 2MASS, unlike {\em IRAS\/} surveys, is mainly
  sensitive to total stellar mass. Consequently, the mass-to-light
  ratio of 2MASS galaxies is expected to have smaller scatter than
  that of {\em IRAS\/} galaxies.
\item Nonlinear bias. Writing Equation~(\ref{eq:grav_bias}) we have
  implicitly assumed that the total flux dipole and the
  magnitude-limited flux dipole differ in the amplitude, but not in
  the direction. However, if large-scale distribution of low-mass DM
  halos is different from the distribution of high-mass halos, then
  the two dipoles will not be collinear.
\item Nonlinear dynamics. The peculiar velocity of the LG is equal to
  the temporal integral of the LG gravitational acceleration along the
  LG trajectory. Therefore, loosely speaking, while the response of
  the LG acceleration to growing nonlinearities is `instantaneous",
  the response of the LG velocity is time-averaged and `retarded'. As
  a result, the acceleration of the LG is more non-linear and higher
  in amplitude (in velocity units) than the velocity of the LG
  (Cieciel{\c a}g et al. 2003). What is more relevant here, at orders
  higher than linear non-local character of gravity reveals itself and
  tends to misalign the velocity vector of the LG with the vector of
  its acceleration. However, the mean misalignment angle between the
  velocity and gravity of the LG-like regions simulated in numerical
  experiments is about $8^\circ$ (Davis \etal 1991, Cieciel{\c a}g
  \etal 2001).

\item Observational effects: shot noise, finite volume of the survey,
  and the mask (due to the zone of avoidance, ZoA). Shot noise and
  finite volume are more an issue for 2MRS, which has a median depth
  of only $60$ \hmpc. Still, we devote Subsection~\ref{sub:shot-noise}
  to a study of shot noise of the 2MASS dipole. M03 perform two
  standard treatments of the masked area: in one of them they clone
  the sky above and below the masked region; in another they fill the
  masked region with randomly chosen galaxies such that it has the
  same surface density as the unmasked area. A recent paper by Tully et al.
  (2008) puts these methods, at least partly, in question. They show
  that there lies a void in the ZoA, which they call the Local Void;
  the LG lies on its boundary. Therefore, in a part of the mask there
  is really nothing, and filling this region with faked galaxies leads
  to a systematic error of the estimate of the LG
  acceleration. However, the role of the Local Void is a recently
  raised issue and we will study it elsewhere.
\end{itemize}

M03 and E06 notice that the misalignment is substantially reduced if
they remove the brightest galaxies in the catalog. M03 remove all
galaxies brighter than $K_s = 8$ mag (375 galaxies), while E06 remove
just five the brightest. They suggest that these galaxies have $M/L
\ne \rm const$ and/or non-linearly contribute to the acceleration. We
will study these issues here. Specifically, the outline of this paper
is as follows. In Section~\ref{sec:descript}, we will present a
formalism which will allow us to model semi-analytically the
distribution function for the misalignment angle between the CMB
dipole and the 2MASS flux dipole. In Section~\ref{sec:nonlin}, we will
model nonlinear effects which appear in such an analysis. The 2MASS
gravity window will be derived and optimized in
Section~\ref{sec:grav}. In Section~\ref{sec:obs} we will account for
observational errors. In Section~\ref{sec:beta} we will present a
formal proof that, under the assumption of negligible shot noise, our
window is indeed optimal. We will also demonstrate how to optimize the
window in presence of shot noise. In Section~\ref{sec:results} we will
show the resulting distribution function for the misalignment angle. A
summary and conclusions will be given in Section~\ref{sec:summ}.

\section{Analytical description of the misalignment}
\label{sec:descript} 
In this Section we will model theoretically the probability
distribution function (PDF) for the misalignment angle between the CMB
dipole and the 2MASS flux dipole. The CMB dipole estimates the
peculiar velocity of the LG, $\bfv$. The 2MASS flux dipole,
Equation~(\ref{eq:grav_bias}), estimates the gravitational
acceleration -- more specifically, the scaled gravity -- of the LG,
$\bfg$, induced by large-scale matter inhomogeneities traced by 2MASS
galaxies.

Let $p(\bfg,\bfv)$ denote the joint PDF for the LG scaled gravity and
peculiar velocity. It is a standard practice to approximate it by a
multivariate Gaussian (S92; S99). Numerical simulations (Kofman \etal
1994, Cieciel{\c a}g \etal 2003) show that nongaussianity of fully
nonlinear $\bfg$ and $\bfv$ is indeed small. This is not surprising
since, e.g.\ gravity is an integral of density over a large volume
(Eq. \ref{eq:clust}), so the central limit theorem can at least partly
be applicable.

Using statistical isotropy of $\bfg$ and $\bfv$, their joint PDF can
be simplified to the form (Juszkiewicz \etal 1990; Lahav, Kaiser \&
Hoffman 1990):
\be p(\bfg,\bfv) = \frac{(1 - \err^2)^{-3/2}}{(2 \pi)^{3}
\sig_\bfg^{3} \sig_\bfv^{3}} \exp\left[- \frac{x^2 + y^2 - 2 \err \mu
x y}{2(1 - \err^2)}\right] \,,
\label{eq:dist}
\ee
where $\sig_\bfg$ and $\sig_\bfv$ are the r.m.s.\ values of a single
Cartesian component of gravity and velocity, respectively. From
isotropy, $\sig_\bfg^2 = \lan \bfg\cdot\bfg \ran/3$ and $\sig_\bfv^2 =
\lan \bfv\cdot\bfv\ran/3$, where $\lan \cdot \ran$ denote the ensemble
averaging. Next, $(\bfx,\bfy) = (\bfg/\sig_\bfg,\bfv/\sig_\bfv)$, and $\mu
= \cos\theta$ with $\theta$ being the misalignment angle between
$\bfg$ and $\bfv$. Finally, $\err$ is the cross-correlation
coefficient of $g_m$ with $v_m$, where $g_m$ ($v_m$) denotes an
arbitrary Cartesian component of $\bfg$ ($\bfv$). From isotropy,

\be
\err = \frac{\lan \bfg \cdot \bfv \ran}{\lan g^2 \ran^{1/2} \lan v^2
\ran^{1/2}} \,.
\label{eq:err}
\ee
Also from isotropy,
\be
\lan x_m y_n \ran = \err \, \de_{mn} \,,
\label{eq:cross}
\ee
where $\de_{mn}$ denotes the Kronecker delta. In other words, there
are no cross-correlations between different spatial components. 

For a given all-sky galaxy survey, the LG gravity is measured
effectively through the window of the survey, $W_\bfg$ (cf.\
Eq.~\ref{eq:clust}):
\begin{equation}
\bfg = \int \frac{{\rm d}^3 r}{4 \pi} \de(\bfr) W_\bfg(\bfr) \frac{\bfr}{r^3}
\,.
\label{eq:gint}
\end{equation}
In contrast, the LG velocity is not estimated from a velocity survey
(i.e., from a catalog of peculiar velocities of galaxies), but
measured directly from the dipole anisotropy of the CMB. Still, to
relate it to theoretical quantities, we write:
\begin{equation}
\bfv = \int \frac{{\rm d}^3 r}{4 \pi} \te(\bfr) W_\bfv(\bfr)
\frac{\bfr}{r^3} \,.
\label{eq:vint}
\end{equation}
Here $\te \equiv - \nabla \cdot \bfv$ is the (minus) velocity
divergence and we assume that the velocity field is
irrotational.\footnote{Kelvin's circulation theorem assures that the
cosmic velocity field is vorticity-free as long as there is no shell
crossing.  N-body simulations (Bertschinger \& Dekel 1989, Mancinelli
et al. 1994, Pichon \& Bernardeau 1999) have shown that the vorticity
of velocity is small in comparison to its divergence even in the fully
nonlinear regime.} Thus, similarly to $\bfg$, $\bfv$ can be expressed
as a Coulomb (Newton) integral over its source, i.e. the field of the
velocity divergence. Here we do {\em not\/} assume that we know the
latter from observations, but we know from theory its statistical
relation to the density field (see this Section and
Section~\ref{sec:nonlin}).  This is sufficient for our purposes in
this work. Since $\bfv$ is directly measured from the CMB dipole, the
effective velocity window, $W_\bfv$, which we have introduced in
Equation~(\ref{eq:vint}), is essentially unity. (Contributions from
all perturbations are included.)\footnote{The velocity of the LG is
fully nonlinear and as such cannot be approximated by low-order
moments of the velocity field. In particular,
$\bfv_{\scriptscriptstyle \rm LG}$ is different from the bulk velocity
of a region around it.} We modify slightly this form of the window to
reflect the finite size of the LG. Following S92 and S99, we adopt
\begin{equation}
W_\bfv = \left\{ \begin{array}{ll}
0, & r < r_{\scriptscriptstyle \rm LG} \,, \\
1, & \mbox{otherwise}\,,
\end{array} \right. 
\label{eq:W_v}
\end{equation}
which has a small-scale cutoff, $r_{\scriptscriptstyle \rm LG} = 1$
\hmpc. This window is markedly different from those appropriate for
velocity surveys: the latter are not spherical, have complicated
shapes and finite depth (Sarkar, Feldman \& Watkins 2007). The gravity
window, $W_\bfg$, of the 2MASS survey is derived in
Section~\ref{sec:grav}.

In Fourier space, relations~(\ref{eq:gint}) and~(\ref{eq:vint}) read:
\begin{equation}
\label{eq:g_Four}
\bfg_{\bfk}=\frac{i\bfk}{k^2}\delta_{\bfk}\widehat{W}_{\bfg}(k),
\end{equation}
\begin{equation}
\label{eq:v_Four}
\bfv_{\bfk}=\frac{i\bfk}{k^2}\te_{\bfk}\widehat{W}_{\bfv}(k),
\end{equation}
where the subscript $\bfk$ denotes the Fourier transform. The quantity
$\widehat{W}$ is related to the window $W$ by the following equation
(S92):
\begin{equation}
\label{eq:W_Four}
\widehat{W}(k) \equiv k \!\int_0^\infty \! W(r) j_1(kr) dr \,.
\end{equation}
Here and below $j_l$ represents the spherical Bessel function of first
kind of order $l$. In particular, 
\be
\widehat{W}_\bfv(k) = j_0(k r_{\scriptscriptstyle \rm LG}) \,.
\label{eq:W_v(k)}
\ee

From equations~(\ref{eq:g_Four}) and~(\ref{eq:v_Four}) we have

\be
\lan \bfg \cdot \bfg \ran = \frac{1}{2 \pi^2} \int_0^\infty
\widehat{W}_\bfg^2(k) P(k) dk \,,
\label{eq:g^2}
\ee
and

\be
\lan \bfv \cdot \bfv \ran = \frac{1}{2 \pi^2} \int_0^\infty
\widehat{W}_\bfv^2(k) P_{\te}(k) dk \,.
\label{eq:v^2_1}
\ee
Here, $P(k)$ and $P_{\te}(k)$ are respectively the power spectrum of
the density and the power spectrum of the velocity divergence. 
Defining 
\be
\calR(k) = \frac{P_{\te}(k)}{P(k)} \,,
\label{eq:calR}  
\ee
we have
\be
\lan \bfv \cdot \bfv \ran = \frac{1}{2 \pi^2} \int_0^\infty
\widehat{W}_\bfv^2(k) \calR(k) P(k) dk \,.
\label{eq:v^2}
\ee
Furthermore,
\be
\lan \bfg \cdot \bfv \ran = \frac{1}{2 \pi^2} \int_0^\infty
\widehat{W}_\bfg(k) \widehat{W}_\bfv(k) C(k) 
P_{\te}^{1/2}(k) P^{1/2}(k) dk ,
\label{eq:g-v}
\ee
where $C(k)$ is the so-called {\em coherence function} (CF), or the
correlation coefficient of the Fourier components of the gravity and
velocity fields (S92):

\be
C(k) \equiv \frac{\lan \bfg_\bfk \cdot \bfv_\bfk^\ast \ran}{\lan
|\bfg_\bfk|^2 \ran^{1/2} \lan |\bfv_\bfk|^2 \ran^{1/2}} 
= \frac{\lan \de_\bfk \te_\bfk^\ast \ran}{\lan |\de_\bfk|^2 
\ran^{1/2} \lan |\te_\bfk|^2 \ran^{1/2}}
\,.
\label{eq:coh_def} 
\ee
Hence, we obtain finally
\be 
\err = \frac{\int_0^\infty \widehat{W}_\bfg(k) \widehat{W}_\bfv(k)
C(k) \calR^{1/2}(k) P(k) dk}{\left[\int_0^\infty \widehat{W}_\bfg^2(k) 
P(k) dk\right]^{1/2} \left[\int_0^\infty \widehat{W}_\bfv^2(k) 
\calR(k) P(k) dk\right]^{1/2}} \,.
\label{eq:err_form} 
\ee 
Equations~(\ref{eq:g^2}), (\ref{eq:v^2}) and~(\ref{eq:err_form})
specify all the parameters (the variances and the correlation
coefficient) that determine the joint PDF for $\bfg$ and $\bfv$,
Equation~(\ref{eq:dist}), {\em in the absence of observational
errors}. The deviation of the correlation coefficient from unity is
then due to different windows, through which the gravity and the
velocity of the LG are measured, and due to nonlinear effects. The
latter are described by two functions: the CF, and the ratio of the
power spectra (Cieciel{\c a}g \& Chodorowski 2004; hereafter C04).

The distribution for the misalignment angle can be derived
from the joint distribution~(\ref{eq:dist}). Here we are interested in
the distribution for the misalignment angle with the observed value of
the LG velocity {\em as a constraint}. The conditional distribution
function, $p(\bfg|\bfv)$, readily results from~(\ref{eq:dist}):
\be
p(\bfg|\bfv) = (2 \pi)^{-3/2} \sig_\bfg^{-3} (1 - \err^2)^{-3/2} 
\exp\left[- \frac{(\bfx - \err \bfy)^2}{2(1 - \err^2)}\right] 
\label{eq:dist_cond}
\ee
(Juszkiewicz \etal 1990; Lahav \etal 1990). The
distribution for the amplitude of the LG acceleration and the cosine
of the misalignment angle is $p(g,\mu|\bfv) = 2 \pi g^2
p(\bfg|\bfv)$. The distribution for $\mu$ is obtained by marginalizing
over $g$,
\be
p(\mu|\bfv) = 2 \pi \! \int_0^\infty {\rm d}g\, g^2 p(\bfg|\bfv) \,,
\label{eq:mu}
\ee
and the distribution for the angle itself is $p(\theta|\bfv) =
|d\mu/d\theta|\, p(\mu|\bfv)  = \sin(\theta)\, p(\mu|\bfv)$. This yields
(Juszkiewicz \etal 1990; Lahav \etal 1990)
\begin{eqnarray}
p(\theta|\bfv) 
\!\!\! &=& \!\!\! \sin(\theta)\, \exp\left(-q^2\right) 
\left\{\frac{q\mu}{\sqrt{\pi}} \right. \nonumber \\
\!\!\! &\hphantom{=}& \!\!\! + \left(\frac{1}{2} + q^2 \mu^2 \right) 
\exp\left(q^2 \mu^2\right) \left[1 + {\rm erf}(q \mu)\right] 
\biggr\} \,,
\label{eq:te|v}
\end{eqnarray}
where
\begin{equation}
q = \frac{\err y}{\sqrt{2(1 - \err^2)}} \,.
\label{eq:q}
\end{equation}
We remind that $y$ is the amplitude of the LG peculiar velocity in
units of the 1D velocity dispersion, $y = v_{\scriptscriptstyle \rm
LG}/\sig_\bfv$. For $v_{\scriptscriptstyle \rm LG} = 627$ \kms\
(Bennett \etal 2003) and the values of the cosmological parameters
adopted here (as described in Section~\ref{sec:nonlin}), $y =
2.64$. This might suggest that the velocity of the LG is a rare event;
however, this is on the contrary. First, one should compare the
amplitude of the LG velocity to the 3D velocity dispersion,
$\sig_{\bfv,\rm 3D} = \sqrt{3} \sig_\bfv$. Therefore, the relevant
parameter here is $y' \equiv v_{\scriptscriptstyle \rm
LG}/\sig_{\bfv,\rm 3D} = 1.52$. Second, the probability that a
randomly chosen region will have velocity greater than
$v_{\scriptscriptstyle \rm LG}$ is equal to $\int_{1.52}^\infty dy'
h(y')$, where $h(y) = \sqrt{2/\pi}\, y^2 {\rm e}^{-y^2/2}$ is the {\em
Maxwellian\/} distribution. For the lower limit of the integral equal
to $2.64$, the value of the integral is $0.07 = 7\%$. However, for
$1.52$, its value is $0.51 = 51\%$.

The misalignment of only several degrees corresponds to a strong
coupling between $\bfg$ and $\bfv$. In the strong coupling limit $1 -
\err \ll 1$, so $q \gg 1$. Also, $\mu$ is then close to unity.
Therefore, in equation~(\ref{eq:te|v}) we can use an asymptotic
formula for the error function,

\be
{\rm erf}(s) \simeq 1 - \frac{1}{\sqrt{\pi} s} {\rm e}^{-s^2} \qquad
\mbox{for } s \gg 1 \,.
\label{eq:erf}
\ee
We then obtain a small-angle approximation of the distribution for the
misalignment angle (Lahav \etal 1990): 
\be
p(\theta|\bfv) \simeq \frac{\theta}{\theta^2_\ast}
\exp\left(- \frac{\theta^2}{2 \theta^2_\ast}\right) , 
\label{eq:te_approx}
\ee
with 
\be 
\theta_\ast = \frac{\sqrt{1 - \err^2}}{\err y} \,.
\label{eq:te_ast}
\ee
Thus, in the strong coupling limit the misalignment angle, given the
velocity constraint, is Rayleigh-distributed. The parameter
$\theta_\ast$, much smaller than unity (in radians), is a
characteristic measure of the misalignment.\footnote{Consequently, we
could approximate $\theta_\ast$ by $\sqrt{2(1 - \err)}/y$. However,
exact expression~(\ref{eq:te_ast}) is not more complex, while it
remains valid also for higher-order corrections to the
distribution~(\ref{eq:te_approx}).} The expectation value of the angle
is
\be
\lan\theta|\bfv\ran = \sqrt{\frac{\pi}{2}} \theta_\ast \,.
\label{eq:exp_te|v}
\ee

Other quantities characterizing the distribution which are of interest
here are quantiles. In our, slightly modified notation, the quantile
$\theta_q$ of a distribution $p(\theta)$ is such a number, that 
\be
\int_{\theta_{\rm min}}^{\theta_q} p(\theta) \, {\rm d}\theta =
\frac{q}{100} \,.
\label{eq:quant_def}
\ee 
For the Rayleigh distribution, ${\theta_{\rm min}} = 0$. For our
purposes, interesting quantiles are
\be
\theta_{10} = \sqrt{2 \ln(10/9)}\, \theta_\ast \,, \quad {\rm and} \quad 
\theta_{90} = \sqrt{2 \ln{10}}\, \theta_\ast \,.
\label{eq:quantiles}
\ee

\section{Nonlinear effects}
\label{sec:nonlin}
Using numerical simulations, C04 modelled the CF
(Eq.~\ref{eq:coh_def}) and the ratio of the power spectra
(Eq.~\ref{eq:calR}). The simulations were evolved from Gaussian
initial conditions. As the initial power spectrum of matter
fluctuations, a cold dark matter (CDM) spectrum was adopted (as in
Eq.~7 of Efstathiou, Bond \& White 1992), with the shape parameter
$\Gamma=0.19$. Both the CF and the ratio of the power spectra were
modelled as functions of the wavevector, $k$, and the amplitude of the
matter fluctuations, $\sigma_8$. For the CF, C04 found the following
fit:

\begin{equation}
C(k) = \left [ 1+(a_0 k - a_2 k^{1.5} + a_1 k^2)^{2.5} \right
]^{-0.2},
\label{eq:CF_fit}
\end{equation}
with the coefficients given by the following, power-law, scaling
relations in $\sigma_8$:
\begin{eqnarray}
a_0 &=& 4.908\ \sigma_8^{0.750} \nonumber ,\\
a_1 &=& 2.663\ \sigma_8^{0.734} , \\
a_2 &=& 5.889\ \sigma_8^{0.714} \nonumber .
\end{eqnarray}
The fit was calculated for $k\in [0, 1]$ $h\; \hbox{Mpc}^{-1}$ and
$\sigma_8 \in [0.1,1]$, with the imposed constraint $C(k=0)=1$. This
constraint assures that for sufficiently large, linear scales, the
relation between the gravity and the velocity is deterministic and
linear (see Eq.~\ref{eq:v-g}). Formula~(\ref{eq:CF_fit}) is a better
fit to the CF than an earlier formula of Chodorowski \& Cieciel{\c a}g
(2002), which was less accurate for low values of $k$. Chodorowski \&
Cieciel{\c a}g (2002) investigated numerically also the dependence of
the CF on $\Omega_m$ and found it to be extremely weak.

Defining the {\em scaled\/} velocity divergence, $\tilde\te \equiv
\Omega_m^{-0.6} \te = - \Omega_m^{-0.6} \nabla \cdot \bfv$, C04 found
the following fit for the ratio of the power spectra:

\be
\calR(k) = [1+(7.071k)^4]^{-\alpha} \,,
\label{eq:pvpgfit_fix}
\ee
where

\be 
\alpha = -0.06574 + 0.29195\sigma_8 \qquad 
\mathrm{for}~0.3<\sigma_8<1 \,.
\ee
C04 argued that the ratio of the power spectra practically does not
depend on the background cosmological model. This ratio is unity in
the linear regime ($k \ll 1$) but decreases in the nonlinear regime,
because the velocity grows slower than it would be expected from the
linear approximation.

In this paper we use Equations~(\ref{eq:CF_fit})
and~(\ref{eq:pvpgfit_fix}) as the formulas respectively for the CF and
the ratio of the power spectra. For $\sigma_8$ we adopt the value
obtained from a joint analysis of third-year WMAP and SDSS, $\sigma_8
= 0.77$ (Spergel \etal 2007). In Equations~(\ref{eq:g^2}),
(\ref{eq:v^2}) and~(\ref{eq:err_form}), as the power spectrum we use a
CDM spectrum. For zero baryon content, the shape parameter of the
spectrum, $\Gamma$, equals simply to $\Omega_m h$. Non-zero baryon
content of the Universe modifies the shape parameter to (Sugiyama
1995) \be \Gamma_{\rm eff} = \Omega_m h \exp\left[- \Omega_b\left(1 +
\sqrt{2 h} / \Omega_m\right) \right] \,.
\label{eq:Gamma_eff}
\ee
Here we adopt $\Gamma_{\rm eff} = 0.15$, the value obtained both from
first-year WMAP (Spergel \etal 2003) and a joint analysis of
third-year WMAP and SDSS (Spergel \etal 2007). This value is in
excellent agreement with the constraint on the shape of the power
spectrum of 2MASS galaxies, $\Gamma_{\rm eff} = 0.14 \pm 0.02$,
obtained by Frith, Outram and Shanks (2005; assuming a flat
$\Lambda$CDM cosmology, a primordial scale-invariant power spectrum
and negligible neutrino mass). It is slightly higher than the
corresponding result of Maller \etal (2005), $\Gamma_{\rm eff} = 0.12
\pm 0.01$, obtained using a measure of the three-dimensional power
spectrum via an inversion of the 2MASS angular correlation function.

\section{Gravity window of 2MASS}
\label{sec:grav}
In this Section we derive the gravity window of the 2MASS survey.  The
2MASS survey is dense, uniform and has an unprecedented sky
coverage. Therefore, to a good accuracy it can be described by a
spherical window. Since distances of 2MASS galaxies are unknown, the
galaxies are weighted only by their fluxes and {\em not\/} by their
distances (like, e.g., by the inverse of the selection function).
This is the central assumption of the calculation below.

The background light intensity due to uniform distribution of discrete
sources is
\begin{equation}
I=\int S\, dN ,
\label{go31}
\end{equation}
where $S = L / (4\pi r^2)$ is the observed flux from the sources with
intrinsic luminosity $L$ and $dN$ is the number of sources per
steradian. In the case of uniformly distributed sources with the
luminosity function (LF) $\Phi(L)$, the contribution from a shell of
thickness $dr$ and radius $r$ is $dN = \Phi(L) \, dL\, r^{2}dr$.
For a flux-, or magnitude-limited survey, only galaxies with $L >
4\pi r^2 S_{\rm min}$ are observed, where $S_{\rm min}$ is the
limiting (minimal) flux. Hence, 
\begin{equation}
I = \int^{\infty}_{0}{\int^{\infty}_{0}{\Theta(r,L) 
\frac{L}{4\pi r^{2}}\,\Phi(L) \, dL \,r^{2}dr}} ,
\label{eq:go35}
\end{equation}
where $\Theta(r,L)$ is the Heaviside step-function, $\Theta_{H}(L-4\pi
r^{2}S_{\rm min})$. Writing $L_{\rm min} \equiv 4\pi r^{2}S_{\rm
min}$, this yields

\begin{equation}
I=\frac{\langle L \rangle N_{0}}{4\pi}\int^{\infty}_{0}{dr\,W_\bfg(r)} .
\label{go39}
\end{equation}
Here, $N_{0} = \int^{\infty}_{0}\Phi(L) \, dL$, and
\be
\langle L \rangle = \frac{\int^{\infty}_{0}{L\Phi(L)
\, dL}}{\int^{\infty}_{0}{\Phi(L) \, dL}}
\label{eq:L_mean}
\ee
is the average luminosity of the population. The flux window of the
survey is
\begin{equation}
W_\bfg(r) = \frac{\int^{\infty}_{L_{\rm min}} 
{L\Phi (L)\, dL}}{\int^{\infty}_{0}{L\Phi (L)\, dL}} .
\label{eq:go38}
\end{equation}
$W_\bfg$ gives the percentage of the total light from distance $r$
which is included in the survey. Loosely speaking, it suppresses
contributions from distances larger than $\sqrt{\langle L \rangle /(4
\pi S_{\rm min})}$. Its detailed form is determined by the LF.

The LF of 2MASS galaxies has been estimated by Bell et al.\ (2003), by
matching a spectroscopic sample of Early Data Release SDSS galaxies
with the 2MASS extended source catalog, to obtain redshifts for a
subsample of 2MASS galaxies. Bell et al.\ fitted the 2MASS LF by the
Schechter function:
\begin{equation}
\Phi(L)\, dL=\Phi^{*} \left( \frac{L}{L^{*}} \right)^\alpha 
\exp \left(-\frac{L}{L^{*}}\right)\frac{dL}{L^{*}} \,,
\label{go34}
\end{equation}
where $\Phi^{*}$ is the LF normalization, $L^{*}$ is the
characteristic luminosity at the `knee' of the LF, where the form
changes from exponential to power law, and $\alpha$ is the `faint end
slope'. For $K_s$-band, they found that $\alpha = -0.77$ and
the absolute magnitude $M^\ast$, corresponding to the absolute
luminosity $L^\ast$, is $M^\ast = -23.29 + 5 \log_{10} h$. We adopt
this form of the 2MASS LF here. 

The flux window of the 2MASS survey can be compared with the selection
function of the survey, defined as

\be
\phi(r) = \frac{\int^{\infty}_{L_{\rm min}} 
{\Phi (L)\, dL}}{\int^{\infty}_{0}{\Phi (L)\, dL}} .
\label{eq:select}
\ee
The selection function gives the probability that a randomly selected
galaxy at distance $r$ will be included in the survey. E06 call the
selection function the `number-weighted selection function', and the
flux window the `flux-weighted (or luminosity-weighted) selection
function'. They note that ``the number-weighted selection function
drops with the distance faster than the luminosity-weighted selection
function. At large distances, we observe only the most luminous
galaxies, so the amount of `missing' luminosity from a volume of space
is not as big as the number of `missing' galaxies''. We fully
agree. This implies in practice that when estimated from a galaxy
survey, the flux dipole is a more robust quantity than the number
dipole. 

If we want to exclude also the brightest sources, then $\Theta(r,L)$
in Equation~(\ref{eq:go35}) becomes the product of two Heaviside
functions, $\Theta(r,L) = \Theta_{H}(L-4\pi r^{2}S_{\rm min})\cdot
\Theta_{H}(4\pi r^{2}S_{\rm max}-L)$. Here, $S_{\rm max}$ is the upper
limiting (maximal) flux. It is simple to check that the survey window
then becomes

\begin{equation}
W_\bfg(r)=\frac{\int^{L_{\rm max}}_{L_{\rm min}} 
{L\Phi (L)\, dL}}{\int^{\infty}_{0}{L\Phi (L)\, dL}} ,
\label{eq:go40}
\end{equation}
where $L_{\rm max} \equiv 4 \pi r^2 S_{\rm max}$.  This windows
suppresses also contributions from distances {\em smaller\/} than
about $\sqrt{\langle L\rangle /(4 \pi S_{\rm max})}$, or, for the
Schechter LF, just about $\sqrt{L^\ast/(4 \pi S_{\rm max})}$.

What would be the gravity window for the {\em number} dipole? The
answer depends on the weighting scheme. In case of the number dipole
all galaxies are weighted equally -- in a sense that they are {\em
not\/} weighted by their fluxes or masses -- but they may, or may not,
be weighted proportionally to the inverse of the selection
function. If they are not, it is clear from the above analysis that
then the gravity window is the selection function, $\phi$. (Now,
instead of missing some percentage of the total light from distance
$r$, we miss some percentage of all galaxies located there.) However,
if they are weighted as $1/\phi(r)$, the gravity window is simply
unity. This is so because the $1/\phi(r)$ weighting corrects for, on
average, missing signal from large distances. The price to pay for
this correction is huge variance of such an estimator of the LG
gravity. This variance is called shot noise (see
Subsection~\ref{sub:shot-noise}) and has dominant contributions from
large scales. Number-, rather than mass-, weighting of galaxies is
another source of variance of this estimator, for simplicity also
called shot noise. That shot noise comes predominantly from small
scales.

Using the {\em IRAS} 1.2 Jy survey, S92 measured the number dipole,
weighting galaxies originally as $1/\phi(r)$. To mitigate shot noise
and nonlinear effects from small scales and shot noise from large
scales, S92 decided to modify these weights. They did this introducing
the so-called standard {\em IRAS} window,
\begin{equation}
W_{\mathit IRAS} = \left\{ \begin{array}{ll} (r/r_s)^3, & r < r_s \,,
\\ 1, & r_s < r < R_{\rm max} \,, \\ 0, & R_{\rm max} < r \,.
\end{array} \right. 
\label{eq:W_g}
\end{equation}
This window is characterized by a small-scale smoothing, $r_s$, and a
sharp large-scale cutoff, $R_{\rm max}$. S99 adopted the values $r_s =
5$ \hmpc\ and $R_{\rm max} = 150$ \hmpc, appropriate for the complete
PSCz catalog (Saunders \etal 2000). Modified weights assigned to {\em
IRAS\/} galaxies by S92 and S99 were
\begin{equation}
{\rm weight}(i) = \frac{W_{\mathit IRAS}(r_i)}{\phi(r_i)} \,.
\label{eq:weight}
\end{equation}
It is clear that in this case, the gravity window of the {\em IRAS\/}
number dipole is $W_{\bfg,\, \mathit IRAS} = W_{\mathit IRAS}$. We
will return to this point later. 

To specify completely the distribution for the misalignment angle
(Eq.~\ref{eq:te|v}, or its small-angle approximation,
Eq.~\ref{eq:te_approx}), we need the value of the velocity variance,
Equation~(\ref{eq:v^2}), and of the correlation coefficient,
Equation~(\ref{eq:err_form}). In order to calculate the latter, we
need to derive the Fourier form (Eq.~\ref{eq:W_Four}) of the 2MASS
window, given above. In Equation~(\ref{eq:W_Four}), we can use the
fact that the spherical Bessel function, $j_1(x) = - (d/dx)j_0(x)$,
and integrate by parts. This yields

\be
\widehat{W}_{\bfg}(k) = \int_0^\infty j_0(kr)\, W_\bfg'(r)\, {\rm d}r .
\label{eq:W_parts}
\ee
Let's cast Equation~(\ref{eq:go40}) to the form

\begin{equation}
W_\bfg(r)=\frac{\int^{L_{\rm max}}_{L_{\rm min}} 
{\Psi(L)\, dL}}{\int^{\infty}_{0}{\Psi (L)\, dL}} ,
\label{eq:W_cast}
\end{equation}
where 

\be
\Psi(L) \equiv L\,\Phi(L) \,.
\label{eq:Psi}
\ee
We can then write
\begin{eqnarray}
W'_{{\bf g}}(r) &=& 
\frac{\partial W_{{\bf g}}(r)}{\partial L_{\rm max}} \frac{dL_{\rm max}}{dr} +
\frac{\partial W_{{\bf g}}(r)}{\partial L_{\rm min}} \frac{dL_{\rm min}}{dr}  
\nonumber \\    &=&
\vbox{\vspace{12pt}} \frac{\Psi (L_{\rm max})\, 8\pi r S_{\rm max} - 
\Psi (L_{\rm min})\, 8\pi r S_{\rm min}}{\int^{\infty}_{0}{\Psi (L)\, dL}} \,.
\label{eq:go47}
\end{eqnarray}
Using Equations~(\ref{eq:W_parts}) and~(\ref{eq:go47}), and the fact
that $j_0(x) = \sin{x}/x$, we obtain
\begin{eqnarray}
\widehat{W}_{{\bf g}}(k) &=&
\frac{8\pi S_{\rm max}}{k \int^{\infty}_{0}{\Psi (L)\, dL}} 
\int^{\infty}_{0}{\sin(kr)\Psi (L_{\rm max}){\rm d}r}\nonumber \\
                         &-& 
\frac{8\pi S_{\rm min}}{k \int^{\infty}_{0}{\Psi (L)\, dL}} 
\int^{\infty}_{0}{\sin(kr)\Psi (L_{\rm min}){\rm d}r} ,
\label{eq:go48}
\end{eqnarray} 
where $\Psi(L)$ is defined by Equation~(\ref{eq:Psi}). 
Note that the flux (or gravity) window does not appear in
Equation~(\ref{eq:v^2}) for the velocity variance, and in
Equation~(\ref{eq:err_form}) for the correlation coefficient it
appears in such a way that its absolute normalization cancels out. In
other words, the PDF for the misalignment angle is sensitive only to
the {\em shape\/} of the gravity window. 

To relate the limiting fluxes to the limiting magnitudes, we remind
that the observed minimal flux $S_{\rm min}$ is related to the
apparent maximal magnitude $K_{\rm max}$ in the following way:

\begin{equation}
S_{\rm min} = S_{0} \,10^{-0.4 K_{\rm max}} , 
\label{eq:go49}
\end{equation}
where $S_0$ is the reference flux, which appears also in the relation
between the absolute magnitude $M^\ast$ and absolute luminosity
$L^\ast$,
\be
M^\ast = -2.5 \log_{10}{\frac{L^\ast}{4\pi (10\, {\rm pc})^2 S_0}} \,.
\label{eq:go50}
\ee
In Equations~(\ref{eq:go49})--(\ref{eq:go50}) we can therefore
eliminate $S_0$, obtaining

\begin{equation}
S_{\rm min} = 1.803 \times 10^{-5} \frac{L^\ast}{4\pi (1 \hmpc)^2} \,.
\label{eq:go51}
\end{equation}
Calculating the numerical coefficient in the above equation we have
adopted $M^\ast = -23.29 + 5 \log_{10} h$ (Bell et al.\
2003). Following M03, for $K_{\rm max}$ we have adopted the value
$13.57$. The reason for this choice of $K_{\rm max}$ is
twofold. First, our aim here is to improve the 2MASS window used by
M03 properly accounting for nonlinear effects, which affect only the
choice of optimal $K_{\rm min}$ (the minimal magnitude). Second, M03
chose $K_{\rm max} = 13.57$ because ``the extended source catalog is
$97.5$\% complete within the SDSS early data release for
extinction-corrected Kron magnitudes of $K_s \le 13.57$ mag'' (Bell et
al.\ 2003, Jarrett 2004). The 2MASS window is the 2MASS flux-weighted selection
function under the assumption that the survey is complete within the
flux limits. If we wanted to go deeper, we should account for
increasing incompleteness as a function of distance. However, Figure 1
of M03, showing the convergence of the 2MASS dipole as a function of
the limiting magnitude, suggests that contributions from {\em all\/}
galaxies (i.e.\ even from those {\em not\/} included in the survey)
fainter than $13.57$ mag are most likely negligible. Even for 2MASS
galaxies {\em brighter\/} than $13.57$ mag (where the catalog is
complete), "the faintest 300,000 galaxies only change the dipole value
by less than 5\%" (M03).

\begin{figure}
  \centering \includegraphics[width=8cm]{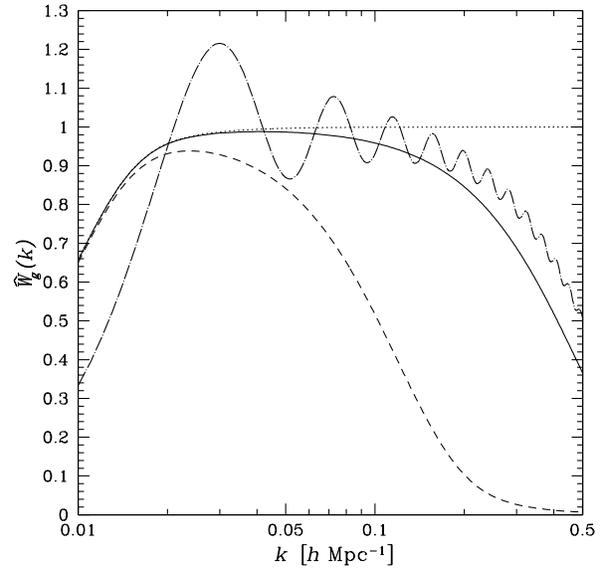} \caption{Gravity
  windows for the 2MASS all-sky survey, for $K_{\rm max} = 13.57~\rm
  mag$ and different values of $K_{\rm min}$. Dotted line corresponds
  to Equation~(\ref{eq:go53}), i.e. to the case where the brightest
  galaxies are not excluded from the calculation of the flux
  dipole. Dashed and solid lines are plotted using
  Equation~(\ref{eq:go48}) and describe the cases of excluding all
  2MASS galaxies brighter than $K_{\rm min} = 8~\rm mag$ and $K_{\rm
  min} = 5~\rm mag$, respectively. Since all 2MASS windows have the
  same $K_{\rm max}$, they are similarly suppressed at large scales,
  i.e., at small $k$. For reference, we also plot the standard {\em
  IRAS\/} window (Eq.~\ref{eq:W_g(k)}), with $r_s = 5$ \hmpc\ and
  $R_{\rm max} = 150$ \hmpc\ (dot-long-dashed line).}
\label{fig:win}
\end{figure}

For $S_{\rm max}$ we have simply
\begin{equation}
S_{\rm max} = S_{\rm min} 10^{0.4 (13.57 - K_{\rm min})} \,.
\label{eq:go52}
\end{equation}
If the brightest galaxies are not excluded, then either directly from
Equation~(\ref{eq:go38}), or from Equation~(\ref{eq:go48}), performing
the limit $S_{\rm max} \to \infty$, we obtain
\begin{equation}
\widehat{W}_{{\bf g}}(k) = 1 -
\frac{8\pi S_{\rm min}}{k \int^{\infty}_{0}{\Psi (L)\, dL}} 
\int^{\infty}_{0}{\sin(kr)\Psi (L_{\rm min}){\rm d}r} .
\label{eq:go53}
\end{equation}
Figure~\ref{fig:win} shows the 2MASS gravity windows for $K_{\rm max}
= 13.57~\rm mag$ and different values of $K_{\rm min}$. Dotted line
corresponds to Equation~(\ref{eq:go53}), i.e. to the case where the
brightest galaxies are not excluded from the calculation of the flux
dipole. Dashed and solid lines are plotted using
Equation~(\ref{eq:go48}) and describe respectively the cases of
excluding all 2MASS galaxies brighter than $K_{\rm min} = 8~\rm mag$
(as done by M03), and $K_{\rm min} = 5~\rm mag$ (our choice, as
justified below). Let's try to understand the influence of the
limiting magnitudes on the shape of the gravity window. Since all
2MASS windows have the same $K_{\rm max}$, they are similarly
suppressed at large scales (small $k$). For a given $K_{\rm min}$
(corresponding to maximal limiting flux), all objects brighter than
$L^{*}$ are excluded from distances smaller than $r_{\rm min} =
\sqrt{L^\ast/(4 \pi S_{\rm max})}$, and at distances $r < r_{\rm
min}$, all sources brighter than $L^{*}(r/r_{\rm min})^2$ are
excluded. For the limiting magnitude $K_{\rm min} = 5$, $r_{\rm min}
\simeq 4.5$ \hmpc, while for $K_{\rm min} = 8$, $r_{\rm min} \simeq
18.1$ \hmpc. Consequently, the window for no exclusion of the
brightest galaxies (dotted line) does not drop down at all for large
$k$ (small scales). The window for $K_{\rm min} = 5$ (solid line) does
drop down but is fairly wide, while the window for $K_{\rm min} = 8$
(dashed line) drops very rapidly. Since, as explained earlier,
$W_{\bfg,\, \mathit IRAS} = W_{\mathit IRAS}$, using
Equation~(\ref{eq:W_g}) we have
\begin{equation}
\widehat{W}_{\bfg,\, \mathit IRAS}(k) = \frac{3 j_1(k r_s)}{k r_s} -
j_0(k R_{\rm max}) \,.
\label{eq:W_g(k)}
\end{equation}
For reference, we plot this standard {\em IRAS\/} window in
Figure~\ref{fig:win} (dot-long-dashed line). The small scale smoothing
of the {\em IRAS\/} window is $r_s = 5$ \hmpc, while for the 2MASS
window with $K_{\rm min} = 5$, the effective smoothing scale $r_{\rm
min}$ is about $4.5$ \hmpc. It is not therefore surprising that at
small scales the {\em IRAS\/} window, except for its oscillatory
behaviour, decreases fairly similarly to this 2MASS window.

Even neglecting shot noise, suppressing contributions to the flux
dipole from small scales is necessary, since nonlinear effects should
be mitigated. For large $k$ the coherence function of velocity with
gravity (Eq.~\ref{eq:CF_fit}) drops significantly below unity,
decreasing the value of the cross-correlation coefficient
(Eq.~\ref{eq:err_form}).  Then suppressing the gravity window for
large $k$ has almost no effect on the cross-term (which is the
numerator of Eq.~\ref{eq:err_form}), while it decreases the gravity
variance, the square root of which appears in the denominator of this
equation. This manipulation on the gravity window helps therefore to
achieve the best possible correlation between the LG velocity and
gravity. However, when one suppresses the gravity window for scales
which are linear enough so that the CF is close to unity, one worsens
the correlation again. This is so because even for linear fields (CF
and the ratio of power spectra equal to unity) the correlation
coefficient decreases for increasingly different windows of velocity
and gravity. As a result, for some value of $r_{\rm min}$, or $K_{\rm
min}$, the correlation coefficient will have a maximum.

\begin{figure}
  \centering
  \includegraphics[width=8cm]{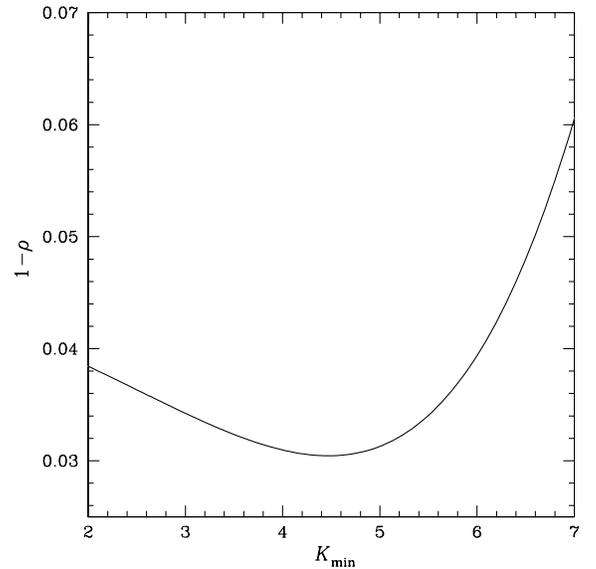}
  \caption{The correlation coefficient, $\err$
   (Eq.~\ref{eq:err_form}), for the 2MASS gravity window
   (Eq.~\ref{eq:go48}), for a range of values of the limiting
   magnitude $K_{\rm min}$. On the ordinate, the value of $1 - \err$
   is plotted. The coefficient has a maximum (respectively, $1 - \err$
   has a minimum) for $K_{\rm min} \simeq 4.5$.}
  \label{fig:err}
\end{figure}

We calculate the correlation coefficient, Equation~(\ref{eq:err_form})
(using the appropriate formulas for the CF and the ratio of the power
spectra, and the velocity window given by Eqs.~\ref{eq:W_v}
and~\ref{eq:W_v(k)}), for the 2MASS gravity window, for a range of
values of the limiting magnitude $K_{\rm min}$. Results are shown in
Figure~\ref{fig:err}. We see that $\err$ has a maximum ($1 - \err$ has
a minimum) for $K_{\rm min} \simeq 4.5$. Either not suppressing small
scales at all ($K_{\rm min} = - \infty$), or suppressing them
excessively ($K_{\rm min}$ greater than, say, 6) clearly decreases the
correlation coefficient. This implies larger statistical errors of the
estimated cosmological parameters when comparing the 2MASS dipole to
the CMB dipole (see Sec.~\ref{sec:beta}); choosing the optimal value
for $K_{\rm min}$ is therefore very important. Instead of $K_{\rm min}
= 4.5$, as the optimal value we have adopted $K_{\rm min} = 5$. We
have done this because the correlation coefficient changes in the
range of $K_{\rm min}$ from 4 to 5 hardly at all, while the number of
excluded galaxies for $K_{\rm min} < 5$ would become very small, resulting in big Poisson noise.

In our analysis so far we have not addressed the effect of shot
noise. If shot noise is not negligible it also increases the optimal
$K_{\rm min}$. This will be explained in Section~\ref{sec:beta}. Shot
noise for the 2MASS flux dipole will be discussed in detail in
Subsection~\ref{sub:shot-noise}.

\section{Observational errors}
\label{sec:obs}
An estimate of the flux dipole from an all-sky galaxy survey is subject to observational errors. One of them is shot-noise, due to
dilute sampling, by distant galaxies, of the underlying mass density
field. Another one is the lack or deficit of galaxies in the Zone of
Avoidance (at low Galactic latitudes). We will describe these errors
correspondingly in Subsections~\ref{sub:shot-noise} and~\ref{sub:mask}.

\subsection{Shot noise}
\label{sub:shot-noise}
The contribution to the LG gravity from a small volume element located
at a distance $r$, estimated from the 2MASS survey is
\begin{equation}
\bfg_E = \frac{\Upsilon}{\varrho_b} {\sum_i}' S_i \hat\bfr_i \simeq 
a \biggl({\sum_i}' L_i \biggr) \hat\bfr \,,
\label{eq:grav_E}
\end{equation} 
where $a = \Upsilon/(4\pi r^2 \varrho_b)$, $\sum_i'$ denotes the sum
over galaxies included in the survey and $\hat\bfr$ is the unit vector
towards the volume element. We {\em model\/} this quantity
theoretically introducing a window function, $W$:
\begin{equation}
\bfg_M = \frac{\Upsilon}{\varrho_b} \sum_i W(r_i) \nu_i S_i \hat\bfr_i
\simeq a W(r) \biggl(\sum_i \nu_i L_i \biggr) \hat\bfr \,.
\label{eq:grav_M}
\end{equation} 
In the above we have accounted for the fact that the mass to light
ratio for an individual galaxy, $\Upsilon_i$, may not be equal to its
average value, $\Upsilon$, but may have some scatter. The quantity
$\nu_i \equiv \Upsilon_i/\Upsilon$; hence $\lan \nu_i \ran = 1$. In
Equation~(\ref{eq:grav_M}) the summation is over all galaxies in the
volume element, regardless whether or not they are in the 2MASS
survey.

The expectation value of the estimated gravity is
\begin{eqnarray}
\lan \bfg_E \ran \!\!\!\! &=& \!\!\!\! a \biggl\lan{\sum_i}' L_i \!
\biggr\ran \hat\bfr = a W_\bfg(r) \biggl\lan{\sum_i} L_i \!
\biggr\ran \hat\bfr \nonumber \\ \!\!\!\! &=& \!\!\!\! a W_\bfg(r)
\lan L \ran N(\Delta V) \hat\bfr \,.
\label{eq:grav_E_mean}
\end{eqnarray}
In the second step we have used the fact that the 2MASS gravity window
we have constructed in Section~\ref{sec:grav}, $W_\bfg$, gives the
percentage of the total light from distance $r$ which is included in
the survey. $N(\Delta V)$ is the number of all galaxies in the volume
element (regardless whether or not included in the survey). The
expectation value of the modelled gravity is
\begin{equation}
\lan \bfg_M \ran = a W(r) \sum_i \lan\nu_i\ran \lan L_i\ran \hat\bfr =
a W(r) \lan L \ran N(\Delta V) \hat\bfr \,.
\label{eq:grav_M_mean}
\end{equation}
In Equation~(\ref{eq:grav_M_mean}) we have assumed that the scatter in
the mass-to-light ratio is independent of luminosity. However, the
above result for $\lan \bfg_M \ran$ is also correct for fairly broad
classes of luminosity-dependent scatter, e.g.\ for $\nu = 1 + \alpha
\calF(L)$, where the luminosity-independent random variable $\alpha$
has zero mean. [In order for the variable $\nu$ to be always positive,
we have to impose an additional constraint $|\alpha_{\rm min}| <
\calF_{\rm max}^{-1}$, where $\alpha_{\rm min}$ (negative since $\lan
\alpha \ran = 0$) is the minimum value of $\alpha$ and $\calF_{\rm
max}$ is the maximum value of the function $\calF$.]

Comparing Equation~(\ref{eq:grav_E_mean}) to~(\ref{eq:grav_M_mean}) we
see that {\em if\/} we adopt $W = W_\bfg$, then the estimated gravity
is an unbiased estimator of the modelled gravity. (From a different
perspective, the modelled gravity is an unbiased estimator of what we
really observe.) Our 2MASS gravity window is constructed precisely in
such a way to assure this. However, the quantity $\bfg_E$ has scatter around $\bfg_M$. For flux (or number) dipoles estimated from
flux-limited galaxy catalogs, the dilute sampling at large distances
introduces significant scatter, called shot noise. The scatter in the
values of $\bfg_E$ around $\bfg_M$ due to the scatter in the
mass-to-light ratio is not, properly speaking, shot noise
(S92). However, for simplicity, we will call both these effects `shot
noise'. (S92 also follow this convention.)

A lucid derivation of shot noise for the number dipole of {\em IRAS\/}
galaxies can be found in Appendix A of S92. Shot noise for the 2MASS
dipole can be calculated in a similar way. However, except for the
fact that the 2MASS dipole is a flux one, there is another important
difference. While in the derivation of S92, galaxies are assigned
weights essentially proportional to the inverse of the selection
function, 2MASS galaxies are given equal weights (their distances are
unknown). We will see below that this introduces a qualitative
difference in the resulting formula for shot noise.

To compute the variance of $\bfg_E$ we take the difference between
$\bfg_E$ and $\bfg_M$ for a full shell of thickness $dr$, we square it
and calculate its expectation value. Finally, we sum up contributions
to the total variance from all shells. The result is:\footnote{In this
paper we need a formula for shot noise only for illustrative purposes,
therefore the derivation will be presented in follow-up work.}
\begin{equation}
\sigma_{\rm SN}^2 = \varrho^{-2}_L {\sum_i}' S_i^2 F(r_i) \!  \left[1
- 2 W_\bfg(r_i) + (1 + Q) \frac{W_\bfg^2(r_i)}{\phi(r_i)}\right] \!\!.
\label{eq:sigma_SN}
\end{equation}
Here, $Q \equiv \lan \nu^2 \ran - 1 \ge 0$ quantifies the amount of
scatter in $M/L$. Were there no scatter, the value of $Q$ would be
zero. The function $F$ is 
\begin{equation}
F(r) = \frac{\int_{L_{\rm min}} L^2 \Phi (L)\, dL \cdot
  \int_{L_{\rm min}} \Phi (L)\, dL }{\left(\int_{L_{\rm min}} L
  \Phi (L)\, dL \right)^2} .
\label{eq:L^2}
\end{equation}
The upper limit in the above integrals is either $L_{\rm max} = 4\pi
r^2 S_{\rm max}$ or infinity, depending on whether we exclude the
brightest objects or not (in the latter case, $S_{\rm max} = \infty$).

Let us now investigate contributions to shot noise from small ($r \to
0$) and large ($r \to \infty$) scales. For $r \to 0$, consider first
the case of no exclusion of the brightest objects. Then, both
$\phi(r)$ and $W_\bfg(r)$ tend to unity. From Equation~(\ref{eq:L^2})
it is obvious that then $F(0)$ is a constant. Therefore,
\begin{equation}
\Delta \sigma_{\rm SN, \: nearby}^2 \propto Q \!\!\! {\sum_{{\rm
nearby,}\, i}\!\!\!\!}' S_i^2 \,.
\label{eq:sigma_SN_nearby}
\end{equation}
If $Q$ is significantly greater than zero, then the RHS of the above
proportionality blows up (since for $r_i \to 0$, $S_i \to
\infty$). This is shot noise from small scales, mentioned already in
Section~\ref{sec:grav}. It similarly plagues the number dipole (see
Eq.~35 of S92, where there are similar $r_i^{-4}$ divergences). As
already mentioned, to mitigate shot noise from small scales S92
introduced a window for the {\em IRAS\/} dipole. With inclusion of the
{\em IRAS\/} window, contributions to shot noise from small scales in
Equation~(35) of S92 are proportional to $W_{\mathit
IRAS}^2(r_i)/r_i^4$. For the standard {\em IRAS\/} window
(Eq.~\ref{eq:W_g}), they scale as $r_i^2 \to 0$ for $r_i \to 0$, so
shot noise from small scales is indeed strongly suppressed. As already
stated, 2MASS galaxies are assigned equal weights. Still, shot noise
from small scales can be mitigated. This is achieved by excluding from
the calculation of the dipole contributions from the brightest
objects, as described below.

For finite $S_{\rm max}$, the selection function is
\begin{equation}
\phi(r) = \frac{\int^{L_{\rm max}}_{L_{\rm min}} 
{\Phi (L)\, dL}}{\int^{\infty}_{0}{\Phi (L)\, dL}} .
\label{eq:select_Lmax}
\end{equation}
Therefore, for $r \to 0$ and finite $S_{\rm max}$, both $W_\bfg(r)$
{\em and\/} $\phi(r)$ tend to zero. However, although $\phi(r)$ tends
to zero, it is straightforward to verify that the quantity $W_\bfg^2(r)
/ \phi(r)$ also tends to zero (at least for the Schechter form of the
luminosity function). Finally, $F(r)$ tends to a constant (though
different from that for the case $S_{\rm max} = \infty$). Hence,
\begin{equation}
\Delta \sigma_{\rm SN, \: nearby}^2 \propto \!\!
{\sum_{\stackrel{\scriptstyle {\rm nearby},\, i}{{\textstyle
\vphantom{M}} S_i < S_{\rm max}}}\!\!\!\!}' S_i^2 \,.
\label{eq:sigma_SN_nearby_Smax}
\end{equation}
The above sum is limited to objects with $S_i < S_{\rm max}$,
what prevents it to blow up. Therefore, excluding the brightest
objects is a good way to mitigate shot noise from small scales having
at one's disposal angular data only.

Contributions to shot noise from large scales do not depend on the
choice whether we exclude the brightest objects, or not. For $r \to
\infty$, both $\phi(r)$ and $W_\bfg(r)$ tend to zero; it is
straightforward to check that then also $W_\bfg^2(r) / \phi(r)$ tends
to zero. The limit of $F(r)$ for $r \to \infty$ is unity. Hence, using
Equation~(\ref{eq:sigma_SN}) we obtain
\begin{equation}
\Delta \sigma_{\rm SN, \: distant}^2 = \varrho^{-2}_L \!\!\!
{\sum_{{\rm distant,}\, i}\!\!\!\!}' S_i^2 \,.
\label{eq:sigma_SN_distant}
\end{equation}
We see that in the case of the flux dipole calculated with equal
weights assigned to all galaxies, shot noise from large scales does
not blow up; on the contrary, it decreases. As a result, in an
analysis of the 2MASS dipole one does not have to exclude any data
from large distances. The analysis presented here assumed $K_{\rm min}
= 13.57$, but it is now clear that when calculating the 2MASS dipole
one can include contributions from 2MASS galaxies fainter than this
magnitude. (Although, as mentioned earlier, Fig.~1 of M03, showing the
convergence of the dipole as a function of the limiting magnitude,
suggests that their contribution will be negligible. See also Fig.~3
of Jarrett 2004.)

Large-scale asymptotic behaviour of shot noise for the flux dipole
(Eq.~\ref{eq:sigma_SN_distant}) with equal weighting is in contrast to
the corresponding behaviour of shot noise for the number dipole
calculated with galaxy weights proportional to the inverse of the
selection function. In the latter case, contributions to shot noise
from large scales diverge as $\phi^{-2}(r_i)$ (see Eq.~35 of S92; at
large distances $\phi(r_i) \ll 1$).\footnote{These divergences are
due to the weighting scheme and not to the type of the dipole.} To
cure this problem, S92 introduced in their standard {\em IRAS\/}
window a sharp large-scale cutoff, $R_{\rm max}$ (see
Eq.~\ref{eq:W_g}). With inclusion of the {\em IRAS\/} window,
contributions from large scales are proportional to $W_{\mathit
IRAS}^2 / \phi^{2}(r_i)$. If $W_{\mathit IRAS}$ is truncated at some
$R_{\rm max}$, then there are no contributions to shot noise from
scales beyond $R_{\rm max}$. Still, this does not imply that
`$1/\phi(r)$' weighting of distant galaxies is a good one. S92 were
aware of this fact and constructed the optimal window for the {\em
IRAS\/} survey, i.e.\ such that it minimized variance of the estimator
of the LG velocity (Eq.~45 of S92). At large scales this window
behaves asymptotically as $J_3(r) \phi(r)$, where $J_3(r) = \int_0^r
{\rm d}^3 r'\, \xi(r')$ and $\xi$ is the mass two-point correlation
function. Therefore, the optimal weighting at large distances is
proportional to $J_3(r)$ (see Eq.~\ref{eq:weight}), so instead of
increasing [as $\phi^{-1}(r)$ does] it decreases to zero, suppressing
shot noise from large scales.  (2MASS weighting is intermediate
between these two extremes.) At small scales, in the absence of the
scatter in the masses of galaxies, the window approaches unity.
Therefore, the $1/\phi(r)$ weighting is then indeed the optimal
one.\footnote{In a related paper, Feldman, Kaiser \& Peacock (1994)
constructed the optimal estimator for the density power spectrum
inferred from redshift surveys. They derived a formula for the optimal
weighting of galaxies (Eq. 2.3.4 of Feldman \etal 1994). For small $r$
the optimal weight behaves like $\phi^{-1}(r)$, while for large $r$ it
approaches asymptotically unity.}  In the presence of scatter the
window filters out small scales, as desired. Surprisingly, S92
resigned from using this window in the analysis of the LG acceleration
and employed instead the standard {\em IRAS\/} window. The reason was
that in the derivation of the optimal window they also attempted to
account for nonlinear effects, but the coherence function they used
was wrong (Chodorowski \& Cieciel\c ag 2002). As a consequence, the
resulting window filtered out small scales excessively. In the {\em
present\/} paper, working with only angular data we have no choice: we
have to assign equal weights to all galaxies. Though this is not the
optimal weighting, this is still quite good: shot noise from large
scales does not blow up. Moreover, excluding the brightest objects
helps to mitigate shot noise from small scales.

Let us recall: we denote the estimated gravity of the LG by $\bfg_E$
(Eq.~\ref{eq:grav_E}) and its modelled gravity by $\bfg_M$
(Eq.~\ref{eq:grav_M}). Although $\bfg_E$ is an unbiased estimator of
$\bfg_M$, it is still a biased estimator of the {\em true\/} gravity
of the LG. Large depth of the 2MASS survey makes the estimated dipole
to converge, but in order to mitigate shot noise and nonlinear effects
from small scales we have to suppress contributions from small
distances. (For angular data the only way to do this is to exclude the
brightest galaxies, located preferentially nearby.) This reduction of
the signal introduces bias in the estimate of the LG gravity. However,
applying a Maximum Likelihood analysis enables one to correct for this
bias and to obtain an unbiased estimate of the parameter $\beta =
\Omega_m^{0.6}/b_L$. This will be discussed in Section~\ref{sec:beta}.

How big is actual shot noise for the 2MASS survey? To answer this
question, M03 performed bootstrap resampling on the 2MASS galaxy
catalog (100 times). They found that the standard deviation of the
dipole direction was a fraction of a degree, and of the dipole
magnitude a fraction of a percent. They concluded that `the systematic
uncertainties are much larger than the shot noise'. Shot noise is
certainly less an issue for the 2MASS dipole than for the {\em IRAS\/}
PSCz dipole (S99) and for the 2MRS dipole (E06). It is smaller for the
2MASS dipole partly due to much bigger number of galaxies in this
survey compared to {\em IRAS\/} PSCz and 2MRS: there are about
$13,000$ galaxies in the PSCz catalog and $23,000$ galaxies in 2MRS,
while for the limiting magnitude $K_s = 13.57$, the 2MASS catalog
contains about 740,000 galaxies (M03). The main reason, however, is
non-weighting of galaxies when calculating the 2MASS dipole. (E06
weighted 2MRS galaxies inversely to the `flux-weighted selection
function', or, in our terminology, the gravity window,
$W_\bfg$). Still, M03 analysed shot noise including all (so also the
brightest) 2MASS galaxies.  Therefore, it is somewhat surprising that
they did not found a trace of shot noise from small scales. A
forthcoming paper of some of us (Bilicki \& Chodorowski, in
preparation) will be devoted to the {\em optimal measurement\/} of the
2MASS dipole. We are planning to reexamine carefully the issue of shot
noise there. Specifically, we are going to repeat the bootstrap
resampling analysis and to compare its results to our analytical
formula for shot noise, Equation~(\ref{eq:sigma_SN}).

At first sight, it may seem surprising that
Equation~(\ref{eq:sigma_SN}) can be used in the case of only angular
data, since radial functions $\phi(r)$, $W_\bfg(r)$ and $F(r)$ appear
in it. However, these functions are uniquely determined by specifying
$K_{\rm max}$ (corresponding to $S_{\rm min}$), $K_{\rm min}$
(corresponding to $S_{\rm max}$) and the luminosity function of the
2MASS galaxies. As described before, this luminosity function has been
estimated e.g.\ by Bell et al. (2003). The only data employed in
Equation~(\ref{eq:sigma_SN}) are fluxes, $S_i$. (One also needs an
estimate of $Q$, quantifying the amount of scatter in $M/L$.) Since we
do not have these data at our disposal yet, for the rest of this paper
we will accept the claim of M03 that shot noise for the 2MASS flux
dipole is negligible.

\subsection{The mask}
\label{sub:mask}
The source of the biggest systematic error in 2MASS remains the lack
or deficit of galaxies in the Zone of Avoidance (at low Galactic
latitudes). M03 masked the region of the ZoA, and repopulated it with
`synthetic galaxies'. In one method they cloned the sky above and below
the masked region. In another method, they filled `the masked region
with randomly chosen galaxies such that it has the same surface
density as the unmasked area'. The first method gave a dipole
pointing towards $l = 263^\circ$, $b = 40^\circ$. The second method
resulted in a dipole pointing towards $l = 266^\circ$, $b =
47^\circ$. M03 adopted the mean of these two measurements as the
best-fit dipole. However, the error bars they attributed to the
mask-filling uncertainty were somewhat underestimated. We will return
to this point later.

The misalignment can be fully represented as a two-dimensional vector
lying on the celestial sphere. In the absence of shot noise, the total
misalignment is a vectorial sum of the cosmologically-originated
misalignment $\bmath{\theta}_c$, described in
Section~\ref{sec:descript}, and the misalignment due to mask,
$\bmath{\theta}_m$:

\be
\bmath{\theta}= \bmath{\theta}_c + \bmath{\theta}_m .
\label{eq:theta_sum}
\ee
We have $\bmath{\theta}_m = (\Delta l,\Delta b)$, where $l$ and $b$
are respectively the Galactic longitude and latitude. Under the
simplest assumption, the distribution function for $\bmath{\theta}_m$
is a bivariate Gaussian of two uncorrelated variables of the same
variance:
\be
p(\Delta l,\Delta b) = (2\pi)^{-1} \sigma^{-2} \exp{\left(- 
\frac{\Delta l^2 + \Delta b^2}{2 \sigma^2}\right)} .
\label{eq:theta_m}
\ee
The distribution for the modulus $\theta_m = \sqrt{\Delta l^2 + \Delta
  b^2}$ results immediately from Equation~(\ref{eq:theta_m}). It is a
  Rayleigh distribution (cf.\ Eq.~\ref{eq:te_approx}),

\be 
p(\theta_m) = \frac{\theta_m}{\sigma^2} \exp\left(- \frac{\theta_m^2}{2
\sigma^2}\right) .
\label{eq:theta_modulus}
\ee
Let us now invert the above reasoning and apply it to the variable
$\theta_c$. Since the distribution for $\theta_c$ is (approximately)
Rayleigh, the distribution for $\bmath{\theta}_c$ is (approximately) a
bivariate Gaussian. The variable $\bmath{\theta}$ is therefore a sum
of two independent bivariate Gaussians, which itself is a bivariate
Gaussian (of uncorrelated variables). Hence, the variable $\theta$ is
Rayleigh-distributed, with the parameter 

\be
\theta_\ast'^{2} = \theta_\ast^2 + \sigma^2 .
\label{eq:theta'}
\ee
Here, $\theta_\ast^2 = \lan\theta_c^2\ran/2$, and $\sigma^2 =
\lan\theta_m^2\ran/2$.

The parameter $\theta_\ast$ is defined by Equation~(\ref{eq:te_ast})
and determined by the LG velocity variance (Eq.~\ref{eq:v^2}) and the
correlation coefficient (Eq.~\ref{eq:err_form}). Let us find an
estimate for the mask variance $\sigma^2$. We have

\be
\hat{\sigma}^2 = \frac{1}{N - 1} \sum_{i=1}^N x_i^2 \,,
\label{eq:sigma_form}
\ee
where $x_i = \sqrt{(l_i - \bar l)^2 + (b_i - \bar b)^2}$, and $(\bar
l,\bar b)$ are the means for the sample. As stated above, M03 study
the effects of two different methods of `repopulating' the masked
regions with galaxies, so $N = 2$. Then $x_2 = x_1$, hence
$\hat{\sigma} = \sqrt{2} x_1$; the factor $\sqrt{2}$ mustn't be
neglected. This yields (in degrees)
\begin{equation}
\hat{\sigma} \simeq 5.4^\circ
\label{eq:sigma_est}
\end{equation} 
(as opposed to $3.4^\circ$, or $4.5^\circ$, finally adopted by M03).

\section{Likelihood for $\bmath{\beta}$}
\label{sec:beta}
We mentioned in Section~\ref{sec:intro} that a comparison between the
CMB dipole and the 2MASS flux dipole (the latter given by
Equation~\ref{eq:vel_fin}) can serve as a method to measure the
parameter $\beta = \Omega_m^{0.6}/b_L$. Of course, it cannot be done
by naive equating of the two dipoles: such an estimate would be
biased. Here we outline a likelihood estimation of $\beta$ (for
details see C04).

In a Bayesian approach, one ascribes {\it a priori} equal
probabilities to values of unknown parameters, which allows us to
express their likelihood function, given $\bfv$ and $\bfg$ of the LG,
via the probability distribution function for $\bfv$ and $\bfg$:
\begin{equation}
\label{eq:Bayes}
\mathcal{L}({\rm param.}) = p(\bfv,\bfg~|~\rm param.) \,.
\end{equation}
As the parameters to be estimated here we adopt $\beta$ and $b_L$; $p$
is given by Equation~(\ref{eq:dist}). Theoretical quantities in this
distribution are $\sig_\bfg$, $\sig_\bfv$, and $\err$. Since now we
account for observational errors, the variance of a {\em single\/}
spatial component of measured gravity, $\sig_\bfg^2$, is a sum of the
1D cosmological component, $\sig_{\bfg,c}^2$, and errors, $\eps^2/3$
($\bmath{\eps}$ denoting 3D errors, including shot noise and the
mask). Here, gravity is inferred from a galaxian, rather than mass,
density field. Therefore, $\sig_{\bfg,c}^2 = b_L^2 \lan g^2 \ran /3$,
where $\lan g^2 \ran = \lan\bfg\cdot\bfg\ran$ is given by
Equation~(\ref{eq:g^2}). To sum up,
\begin{equation}
\sig_\bfg^2 = \frac{b_L^2 \lan g^2 \ran + \eps^2}{3} \,.
\label{eq:sigma_g}
\end{equation}

Errors in the measured velocity of the LG are negligible compared to
those in the gravity. The relation between the physical velocity,
$\bfv_{\rm ph}$, and the {\em scaled\/} velocity used in this paper,
$\bfv$, is $\bfv_{\rm ph} = \Omega_m^{0.6} \bfv$, hence 1D velocity
variance is 
\begin{equation} 
\sig_\bfv^2 = \frac{\Omega_m^{1.2} s_\bfv^2}{3} = 
\frac{\beta^2 b_L^2 s_\bfv^2}{3} \,,
\label{eq:sigma_v}
\end{equation}
where $s_\bfv^2 \equiv \lan v^2 \ran = \lan\bfv\cdot\bfv\ran$ is given
by Equation~(\ref{eq:v^2}). Finally, errors in the estimate of the LG
gravity do not affect the cross-correlation between the LG gravity and
velocity, but increase the gravity variance. This has the effect of
lowering the value of the cross-correlation coefficient. Specifically,
\begin{equation}
\err' = \err \left(1+ \frac{\eps^2}{b_L^2 \lan g^2 \ran}
\right)^{\! -1/2} ,
\label{eq:err_errors} 
\end{equation}
where $\err$ is given by Equation~(\ref{eq:err_form}). 

From Equation~(\ref{eq:dist}), the logarithmic likelihood for $\beta$
and $b_L$ takes the form:
\begin{eqnarray}
\ln\mathcal{L} (\!\!\!\!\!\!\! &\beta& \!\!\!\!\!\!\!,b_L) = - 3
\ln{(2\pi)} - 3 \ln\left[\sig_\bfg b_L s_\bfv 
(1-\err'^2)^{1/2}\right] - 3 \ln\beta \nonumber \\ 
&-& \!\!\!\! \frac{1}{2(1-\err'^2)} \left( \frac{g_{\rm
m}^2}{\sig_\bfg^2} + \frac{3 v_{\rm m}^2}{\beta^2 b_L^2 s_\bfv^2} -
\frac{2\sqrt{3} \err'\mu_{\rm m} g_{\rm m} v_{\rm m}}{\sig_\bfg \beta
b_L s_\bfv} \right) . 
\label{eq:likeli}
\end{eqnarray}
In the above likelihood, the `data' are the measured values of the LG
gravity and velocity, $g_{\rm m}$ and $v_{\rm m}$, respectively, as
well as $\mu_{\rm m}$, i.e.\ cosine of the misalignment angle. The
model parameters $\sig_\bfg$ and $\err'$ depend solely on $b_L$;
$s_\bfv$ depends neither on $\beta$ nor on $b_L$.

We have written down the expression for the likelihood only for
illustrative purposes. Therefore, for simplicity we will restrict our
analysis to the case of given $b_L$. Then, to find a maximum of the
likelihood we calculate its partial derivative with respect to $\beta$
and equate it to zero. This yields the following equation:
\begin{equation}
\label{eq:kwadrat}
3(1-\err'^2)\beta^2 + \frac{\sqrt{3}\err'\mu_{\rm m} g_{\rm m} v_{\rm
m}}{\sig_\bfg b_L s_\bfv} \beta - \frac{3 v_{\rm m}^2}{b_L^2 s_\bfv^2}
= 0 \,.
\end{equation}
The LG gravity, inferred from the 2MASS survey, is tightly coupled to
its velocity: $1 - \err' \ll 1$ and $1 - \mu_{\rm m} \ll 1$. (See
Table~\ref{tab:param}; $\theta_{\rm obs} = 16^\circ$ corresponds to
$\mu_{\rm m} = 0.96$). At first approximation we can therefore assume
$\err' = \mu_{\rm m} = 1$, hence
\begin{equation}
\label{eq:beta}
\hat{\beta} \simeq \sqrt{3} \frac{\sig_\bfg}{b_L s_\bfv} \frac{v_{\rm
m}}{g_{\rm m}} = \left(\frac{b_L^2 \lan g^2 \ran + \eps^2}{b_L^2 \lan
v^2 \ran}\right)^{1/2} \frac{v_{\rm m}}{g_{\rm m}} \,.
\end{equation}
Thus, the estimate of $\beta$ is not just the ratio of the LG velocity
to its gravity: it is modified by nonlinear effects (which affect
$\lan v^2 \ran$ through the function $\calR$), different observational
windows (which affect differently $\lan g^2 \ran$ and $\lan v^2
\ran$), and observational errors. If all these factors are properly
accounted for, then the estimate of $\beta$ is unbiased.

An optimal estimator is such that is not only unbiased but also has
minimal variance. Expanding the logarithmic likelihood
(Eq.~\ref{eq:likeli}) around its maximum up to second order in $\beta$
enables one to find an estimator of the variance of $\beta$. In the
strong-coupling regime ($1 - \err' \ll 1$, $1 - \mu_{\rm m} \ll 1$),
it is
\begin{eqnarray}
\hat{\sig_\beta^2} \!\!\!\! &=& \!\!\!\!  \frac{v_{\rm m}^2}{\lan v^2
\ran} \left(\frac{b_L^2 \lan g^2 \ran + \eps^2}{g_{\rm m}^2}
\right)^{\! 2} \left(1 - \err'^2\right) \nonumber \\ 
\!\!\!\! &=& \!\!\!\!  
\vbox{\vspace{20pt}} \frac{v_{\rm m}^2 \left(b_L^2 \lan g^2 \ran +
\eps^2\right)}{\lan v^2 \ran\, g_{\rm m}^{4}} \left[b_L^2 \lan g^2
\ran \left(1 - \err^2\right) + \eps^2\right] \,.
\label{eq:var_beta}
\end{eqnarray}
If errors are constant, i.e. they do not depend on $K_{\rm min}$, then
a minimum of the variance corresponds to a maximum of the
cross-correlation coefficient $\err$. (The dependence of $\lan g^2
\ran$ on $K_{\rm min}$ is very weak.) Including higher-order
corrections to the above formula does not change this fact. In the
{\em present\/} paper, errors are indeed constant: the error due to
the mask obviously does not depend on $K_{\rm min}$ and shot noise is
assumed to be negligible. The window function of the 2MASS survey we
have constructed here maximizes $\err$ (see Fig.~\ref{fig:err}). This
is why we call this window, under the assumption of negligible shot
noise, optimal. It exactly corresponds to the minimal expectation
value of the misalignment angle.

As mentioned earlier, in follow-up work we will estimate shot noise
ourselves. If we find that it is in fact {\em not\/} negligible, then
it will influence the optimal value of $K_{\rm min}$. Shot noise as a
function of $K_{\rm min}$ monotonically decreases (see
Eq.~\ref{eq:sigma_SN_nearby_Smax}). The factor $1 - \err^2$, starting
from the value of $K_{\rm min}$ which maximizes $\err$, monotonically
increases (see Fig.~\ref{fig:err}). The interplay between these two
opposing effects in Equation~(\ref{eq:var_beta}) shifts the optimal
$K_{\rm min}$ (corresponding to a minimum of the variance of the
estimator of $\beta$) to a larger value, compared to the case of
negligible shot noise.

\section{Resulting distributions for the misalignment}
\label{sec:results}
\begin{figure}
  \includegraphics[width=8cm]{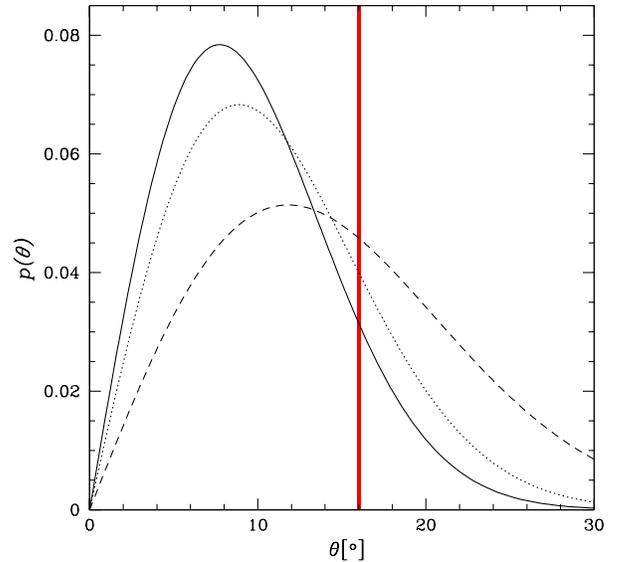}
  \caption{Probability distribution functions for the misalignment
  angle between the 2MASS and CMB dipoles, for various forms of the
  2MASS gravity window. Dotted line is for the case where the
  brightest galaxies are not excluded from the calculation of the flux
  dipole. Dashed line describes the case of excluding all 2MASS
  galaxies brighter than $K_{\rm min} = 8~\rm mag$; solid line
  describes the case of excluding all 2MASS galaxies brighter than
  $K_{\rm min} = 5~\rm mag$. Thick solid vertical line shows the
  measured value of the misalignment between the CMB dipole and the
  2MASS flux dipole as calculated by M03, including {\em all\/}
  galaxies brighter than $K_s = 13.57$ ($16^\circ$).}
\label{fig:PDFs}
\end{figure}

Figure~\ref{fig:PDFs} shows the resulting PDFs for the misalignment
angle, for various forms of the 2MASS gravity window. Like previously,
dotted line corresponds to the case where the brightest galaxies are
not excluded from the calculation of the flux dipole. The vertical stripe shows the value of the misalignment between the CMB
dipole and the 2MASS flux dipole as calculated by M03, including {\em
all\/} galaxies brighter than $K_s = 13.57$. The `observed' value
($16^\circ$) is greater than the expectation value for the angle
($11.1^\circ$), but smaller than $\theta_{90} = 19.0^\circ$ (see
Table~\ref{tab:param}).

\begin{table}
\caption{Parameters of the distribution for the misalignment angle
  between the 2MASS and CMB dipoles, for various forms of the 2MASS
  gravity window. First column shows the limiting magnitude of the
  excluded brightest galaxies, $K_{\rm min}$. Second column shows the
  correlation coefficient of the LG velocity and gravity, $\rho$,
  calculated according to Equation~(\ref{eq:err_form}). Third column
  shows the characteristic value of the misalignment angle (in
  radians), $\theta_\ast$, calculated using
  Equation~(\ref{eq:te_ast}). Fourth column shows the corresponding
  value of the misalignment angle (in radians), including
  observational errors due to the mask, $\theta_\ast'$, calculated
  using Equation~(\ref{eq:theta'}). Fifth column shows the
  expectation value of the misalignment angle (in degrees), $\langle
  \theta\rangle$. Sixth and seventh columns show,
  respectively, the quantiles $\theta_{10}$ and $\theta_{90}$ (in degrees),
  defining the confidence intervals of $10$ and $90$\% (for details
  see text). The last column shows the observed values of the
  misalignment angle (in degrees), obtained with and without exclusion
  of the brightest galaxies.
\label{tab:param}}
\begin{center}
\begin{tabular}{|c|c|c|c|c|c|c|c|}
\hline
\rule{0pt}{2.5ex}
$\!\!\!\!\! K_{\rm min} \!\!\!\!\!$ & $\err$ & $\theta_\ast$ & $\theta_\ast'$ 
& $\!\langle \theta\rangle \,[^\circ]$ & $\theta_{10} \,[^\circ]\!$ 
& $\!\theta_{90} \,[^\circ]\!$ & $\!\theta_{\rm obs} \,[^\circ] \!\!\!$ \\
\hline \hline 
\rule{0pt}{2.5ex}
$\!\!\!\!\!$ -- $\!\!\!\!\!$ & 0.951  & 0.123 & 0.155 & $11.1$ & 4.1 & 19.0 & 16.0 \\
\hline
\rule{0pt}{2.5ex}
$\!\!\!\!\! 8 \!\!\!\!\!$ & 0.901 & 0.183 & 0.206 & $14.8$ & 5.4 & 25.3 & \hspace{0.8ex} 5.2 \\
\hline
\rule{0pt}{2.5ex}
$\!\!\!\!\! 5 \!\!\!\!\!$ & 0.969  & 0.097 & 0.135 & \hspace{0.8ex} $9.7$ & 3.6 & 16.6 & 
\hspace{0.3ex} --- \\
\hline
\end{tabular}
\end{center}
\end{table}

To decrease the misalignment, in the second step M03 excluded from the
analysis all galaxies brighter than $K_{\rm min} = 8$ mag. A dashed
line is plotted for the gravity window corresponding to this
case. Consistently with Figure~\ref{fig:err}, the expectation value
for the angle does not decrease; on the contrary, it increases to
$14.8^\circ$ (Table~\ref{tab:param}). Consequently, one would then
expect the misalignment rather to increase. However, M03 noticed a
substantial decrease of the misalignment, to about $5.2^\circ$. This
value is smaller than the corresponding $\theta_{10} = 5.4^\circ$.
Therefore, there is less than 10\% chance that the decrease might have
been accidental. Rather, an error in the analysis is more likely.
Indeed, E06 repeated the procedure of M03 for the 2MRS data and
essentially did not observe the decrease of the misalignment. 2MRS
survey misses faint galaxies (fainter than $K_s = 11.25$), but does
not miss bright galaxies. Therefore, if the effect was real, one
should observe it also when using the 2MRS data.

Solid line in Figure~\ref{fig:PDFs} is plotted using the window
excluding 2MASS galaxies brighter than $K_{\rm min} = 5~\rm mag$. As
described in Section~\ref{sec:grav}, we expect this window to be close
to optimal. Indeed, the resulting distribution is the narrowest among
the three plotted; the expectation value of the misalignment drops to
$9.7^\circ$ and $\theta_{90} = 16.6^\circ$ (Table~\ref{tab:param}). 
Therefore, with (almost) 90\% confidence we can expect the angle to
decrease when performing such a preselection on 2MASS galaxies. Of
course, this assumes constant mass-to-light ratio for all remaining
(i.e., included) galaxies. 

The window with $K_{\rm min} = 8~\rm mag$ is not optimal because it
excessively mitigates nonlinear effects. This window excludes
too many galaxies: while the window with $K_{\rm min} = 5~\rm mag$
excludes all $L_\ast$ and brighter galaxies closer to the LG than
about $5$ \hmpc, the window with $K_{\rm min} = 8~\rm mag$ does the
same for the distance of about $18$ \hmpc\ (see
Sec.~\ref{sec:grav}). The signal from scales $5$--$18$ \hmpc\ is
sufficiently `linear' to increase (if included) the correlation
between the LG velocity and its measured gravity.

M03 found 375 2MASS galaxies brighter than $K_{\rm min} = 8~\rm
mag$. Based on this number and the relation $N_{\rm excl} \propto
S_{\rm max}^{-3/2}$, where $N_{\rm excl}$ is the number of excluded
galaxies, we predict about six 2MASS galaxies to be brighter than
$K_{\rm min} = 5~\rm mag$. To reduce the misalignment calculated using
their sample, E06 excluded five the brightest galaxies in 2MRS. They
noticed a significant decrease of the misalignment, from about
$21^\circ$ to $14^\circ$. The five most luminous galaxies in 2MRS are
also the five most luminous galaxies in 2MASS. Therefore, exclusion of
these galaxies should work also for denser and deeper 2MASS survey.

\section{Summary}
\label{sec:summ}

\begin{itemize}
\item An ultimate goal of comparing the CMB dipole to the 2MASS dipole
is an estimation of the cosmological parameter $\beta \equiv
\Omega_m^{0.6}/b_L$.

\item To obtain an unbiased estimate of $\beta$, a good and standard
  method is Maximum Likelihood.

\item An important ingredient of this Likelihood analysis is
  the observational window through which the 2MASS flux dipole is
  measured, called here the gravity window of 2MASS. This window
  should be properly modelled.

\item By definition, the optimal window minimizes variance in the
  estimate of $\beta$; optimizing the 2MASS window is therefore
  important. 

\item In this paper, we have modelled the 2MASS gravity window and
  optimized it under the assumption of negligible shot noise. This
  optimization has been achieved by excluding contributions to the
  dipole from the brightest galaxies (Sec.~\ref{sec:grav}). Such an
  exclusion mitigates nonlinear effects from small scales, which
  decorrelate the LG velocity from the estimated LG gravity. We have
  found that the optimal value of the minimal limiting magnitude,
  $K_{\rm min}$ (corresponding to maximal limiting flux), is about
  5.

\item We have also demonstrated how to optimize the window in presence
  of shot noise. We have shown that the optimal value of $K_{\rm min}$
  will increase. 

\item The misalignment angle is a sensitive measure of the correlation
between the two dipoles: the higher the correlation, the smaller the
expectation value of the angle
(Eqs.~\ref{eq:te_ast}--\ref{eq:exp_te|v}). We have shown that a
minimum of the misalignment corresponds to minimal variance of the
estimator of $\beta$. A minimum of the misalignment is thus a sign
of the optimal gravity window.

\item We have modelled analytically the probability distribution
  function for the misalignment angle (Sec.~\ref{sec:descript},
  App.~\ref{app:beyond}). We have shown that the misalignment
  estimated by M03 is consistent with the assumed underlying model
  (though it is greater than the expectation value). We have predicted
  that the misalignment is likely to decrease if 2MASS galaxies
  brighter than $K_{\rm min} = 5~\rm mag$ are excluded from the
  calculation of the flux dipole. This prediction has been indirectly
  confirmed by the results of E06.

\item In a future work, we plan to perform the {\em optimal\/}
  measurement of the value of $\beta$ by comparing the CMB dipole to
  the 2MASS dipole. We will thus have to fully specify the optimal
  window in presence of shot noise (though M03 claim that shot noise
  of 2MASS survey is negligible). An estimate of shot noise can be
  obtained using methods described in
  Subsection~\ref{sub:shot-noise}. However, the misalignment angle can
  be used as an alternative way of optimizing the window. As a
  function of $K_{\rm min}$, the measured value of the misalignment
  will -- with some scatter -- initially decrease, reach a minimum and
  then increase (see Fig.~\ref{fig:err}). It is now clear that the
  value of $K_{\rm min}$ for which the measured misalignment has a
  minimum will be close to that optimizing the measurement of $\beta$.

\end{itemize}

\section*{Acknowledgments}
This work was carried out within the framework of the PAN/CNRS
European Associated Laboratory (LEA) `Astrophysics Poland--France'. It
was also partially supported by the Polish Ministry of Science and
Higher Education under grant N N203 0253 33, allocated for the period
2007--2010.

\appendix
\section{Beyond the small-angle limit}
\label{app:beyond} 
Here we check the accuracy of the small-angle approximation,
Equation~(\ref{eq:te_approx}), of the distribution function for the
misalignment angle, given in general by Equation~(\ref{eq:te|v}). First,
integrating by parts one can show that for $\nu \gg 1$,

\be
\int_\nu^\infty {\rm e}^{-\nu'^2/2} {\rm d} \nu' = \left[\frac{1}{\nu} -
\frac{1}{\nu^3} + \calO\left(\nu^{-5}\right) \right] {\rm e}^{-\nu^2/2} \,.
\label{eq:asympt}
\ee
This yields for the error function a higher-order expansion (than
Eq.~\ref{eq:erf}):
\be 
{\rm erf}(s) \simeq 1 - \left(\frac{1}{\sqrt{\pi} s} - \frac{1}{2
\sqrt{\pi} s^3}\right) {\rm e}^{-s^2}, \qquad s \gg 1 \,.  
\label{eq:erf2} 
\ee
Using this expansion in Equation~(\ref{eq:te|v}) for $\mu > 0$, we obtain

\be
p(\theta|\bfv) = \sin\theta \left(1 + \frac{\cos^2\!\theta}{\theta^2_\ast}
\right) \exp\left(- \frac{\sin^2\!\theta}{2 \theta^2_\ast}\right) , 
\label{eq:te_approx2}
\ee
where $\theta_\ast$ is given by Equation~(\ref{eq:te_ast}) and $\theta
< \pi/2$. For $\theta_\ast\! \to 0$, this distribution simplifies to the
Rayleigh distribution (Eq.~\ref{eq:te_approx}), as expected. 

For $\mu < 0$, one can show in a similar way that
\be 
{\rm erf}(q\mu) \simeq -1 + \left(\frac{1}{\sqrt{\pi} q|\mu|} - \frac{1}{2
\sqrt{\pi} q^3|\mu|^3}\right) {\rm e}^{-q^2\mu^2} \!, \: q \gg 1 .
\label{eq:erf3} 
\ee
Using the latter expansion in Equation~(\ref{eq:te|v}) yields
\be
p(\theta|\bfv) \equiv 0 \qquad {\rm for}\:\: \pi/2 < \theta \le \pi \,.
\label{eq:te_approx3}
\ee
Density distribution~(\ref{eq:te_approx2})--(\ref{eq:te_approx3}) has
analytical {\em cumulative\/} distribution function:
\be
F(\theta) = \left\{ \begin{array}{ll}
1 - \cos\theta\, \exp\left(- \frac{\sin^2\!\theta}{2
		    \theta^2_\ast}\right), & \theta \le \pi/2 \,, \\
1 \,, & \pi/2 < \theta \le \pi \,. \end{array} \right. 
\label{eq:distribuant}
\ee
This allows for a straightforward estimation of the quantiles. Moreover,
using the fact that $p = dF/d\theta$ and integrating by parts, the
expectation value of the angle can be readily calculated:
\be
\lan \theta|\bfv \ran = \sqrt{\frac{\pi}{2}} \theta_\ast\, {\rm erf}
\left(\theta_\ast^{-1}\right) \simeq \sqrt{\frac{\pi}{2}} \theta_\ast - 
\calO \left(\theta_\ast^2 {\rm e}^{-1/\theta_\ast^2}\right) \,.
\label{eq:mean_te_approx2} 
\ee 
Thus, for $\theta_\ast^2 \ll 1$,
distribution~(\ref{eq:te_approx2})--(\ref{eq:te_approx3}) has (almost)
identical mean to the Rayleigh distribution (Eq.~\ref{eq:exp_te|v}).

In practice, the total misalignment angle is a convolution of the
cosmologically-originated misalignment, $\bmath{\theta}_c$, and the
misalignment due to mask, $\bmath{\theta}_m$. This convolution is not as
simple as when both $\bmath{\theta}_m$ {\em and\/} $\bmath{\theta}_c$
are bivariate Gaussians. However, non-Gaussianity of $\bmath{\theta}_c +
\bmath{\theta}_m$ is smaller than of the variable $\bmath{\theta}_c'$
with $\theta_\ast' = \sqrt{\theta_\ast^2 + \sigma^2}$ (because
$\bmath{\theta}_m$ {\em is\/} Gaussian). We see in Table~\ref{tab:param}
that $\theta_\ast'$ is at most $0.2$. For $\theta_\ast' \lta 0.2$,
both distribution~(\ref{eq:te_approx2})--(\ref{eq:te_approx3}) {\em
and\/} the Rayleigh distribution approximate the exact one
(Eq.~\ref{eq:te|v}) very well. Specifically, since then
$\theta_\ast'^2 \ll 1$, we can expand
distribution~(\ref{eq:te_approx2})--(\ref{eq:te_approx3}), obtaining

\be
p(\theta|\bfv) \simeq M(\theta) \frac{\theta}{\theta_\ast'^2}
\exp\left(- \frac{\theta^2}{2 \theta_\ast'^2}\right) ,
\label{eq:te_approx_exp}
\ee
where 

\be
M(\theta) = 1 + \theta_\ast'^2 - \frac{7}{6} \theta^2 +
\frac{\theta^4}{6 \theta_\ast'^2} + \calO \left(\theta_\ast'^4 \right) .
\label{eq:modif}
\ee
It is straightforward to check that approximate
distribution~(\ref{eq:te_approx_exp}) is properly normalized. A
simple calculation yields
\be
\left\lan \theta^2|\bfv \right\ran = 2 \theta_\ast'^2 \left( 1 + 
\frac{\theta_\ast'^2}{3} \right) + 
\calO \left(\theta_\ast'^6 \right) .
\label{eq:te^2_approx}
\ee
Hence, for $\theta_\ast' = 0.2$, the Rayleigh distribution approximates
the second moment of distribution~(\ref{eq:te_approx_exp}) (which, in
turn, is then an excellent approximation of the exact distribution) to
$1.3$\% accuracy. Similarly simple calculations can be performed for
other even moments.

Summing up, the distribution function for the misalignment angle between
the CMB and 2MASS dipoles can be very well approximated by its
small-angle limit, Rayleigh distribution. It may be worth noting for
other applications that, while for $\theta_\ast' > 0.2$ the exact
distribution starts to deviate from the Rayleigh form, it is still well
approximated by
distribution~(\ref{eq:te_approx2})--(\ref{eq:te_approx3}), up to
$\theta_\ast' \sim 0.5$ (except for the very tail). In particular, the
values of the quantiles $\theta_{10}$ and $\theta_{90}$ and of the mean
angle remain within 2\% from the exact values.

\end{document}